\documentclass[fleqn,usenatbib]{mnras}

\usepackage{newtxtext,newtxmath}
\usepackage{breqn}

\usepackage[T1]{fontenc}
\usepackage{ae,aecompl}

\usepackage{graphicx}	
\usepackage{amsmath}	
\usepackage{amssymb}	
\usepackage{xcolor}

\title[Mitigating Biases in Emission Spectra]{Understanding and Mitigating Biases when Studying Inhomogeneous Emission Spectra with JWST}

\author[Taylor et al.]{
Jake Taylor$^{1}$\thanks{E-mail: jake.taylor@physics.ox.ac.uk},
Vivien Parmentier$^{1}$,
Patrick G.J. Irwin$^{1}$,
Suzanne Aigrain$^{2}$,
\newauthor
Elspeth K. H. Lee$^{1}$ \& Joshua Krissansen-Totton$^{3}$
\\
$^{1}$Department of Physics (Atmospheric, Oceanic and Planetary Physics), University of Oxford, Parks Rd, Oxford, OX1 3PU, UK\\
$^{2}$Department of Physics (Astrophysics), University of Oxford,
Denys Wilkinson Building, Keble Rd,
Oxford, OX1 3RH, UK \\
$^{3}$Department of Earth and Space Sciences/Astrobiology Program, Johnson Hall, University of Washington, Seattle, WA, 98195, USA
}

\date{Accepted XXX. Received YYY; in original form ZZZ}

\pubyear{2020}

\begin{document}
\label{firstpage}
\pagerange{\pageref{firstpage}--\pageref{lastpage}}
\maketitle

\begin{abstract}
Exoplanet emission spectra are often modelled assuming that the hemisphere observed is well represented by a horizontally homogenised atmosphere. However this approximation will likely fail for planets with a large temperature contrast in the James Webb Space Telescope (JWST) era, potentially leading to erroneous interpretations of spectra.

We first develop an analytic formulation to quantify the signal-to-noise ratio and wavelength coverage necessary to disentangle temperature inhomogeneities from a hemispherically averaged spectrum. We find that for a given signal-to-noise ratio, observations at shorter wavelengths are better at detecting the presence of inhomogeneities.
We then determine why the presence of an inhomogeneous thermal structure can lead to spurious molecular detections when assuming a fully homogenised planet in the retrieval process. 

Finally, we quantify more precisely the potential biases by modelling a suite of hot Jupiter spectra, varying the spatial contributions of a hot and a cold region, as would be observed by the different instruments of JWST/NIRSpec. We then retrieve the abundances and temperature profiles from the synthetic observations.
We find that in most cases, assuming a homogeneous thermal structure when retrieving the atmospheric chemistry leads to biased results, and spurious molecular detection.
Explicitly modelling the data using two profiles avoids these biases, and is statistically supported provided the wavelength coverage is wide enough, and crucially also spanning shorter wavelengths. For the high contrast used here, a single profile with a dilution factor performs as well as the two-profile case, with only one additional parameter compared to the 1-D approach.


\end{abstract}




\section{Introduction}
 
Exoplanet atmospheres have been studied through transmission and emission spectroscopy in the past 15 years. Since the first detection by~\citep{charbonneau2005detection}, ground and space based observatories have provided a wealth of observations. Emission spectra have mostly been carried observed with the Hubble Space Telescope (HST) and the Spitzer Space Telescope, revealing a diversity of thermal structure and atmospheric composition~\citep[e.g.][]{kreidberg2014precise,stevenson2017spitzer,schwartz2017phase,parmentier2018thermal,irwin20192pt5D,evans2019,garhart2019statistical}. However, most of these observations cover a narrow wavelength range (1.1-1.7$\mu$m for HST) or are only band averaged measurements. In the coming decades, new telescopes such as the James Webb Space Telescope (JWST)~\citep{greene2016characterizing}, the Atmospheric Remote-sensing Infrared Exoplanet Large-survey (ARIEL)~\citep{tinetti2018chemical} and the E-ELT \citep{brandl2018status} will improve both the precision and the wavelength coverage of these observations by an order of magnitude.
 
 


One method of studying the atmospheric constituents of an exoplanet is by employing an atmospheric retrieval~\citep{madhusudhan2009temperature,line2013systematic,waldmann2015tau,waldmann2015rex,blecic2017implications,cubillos2017aerosol,petitradtran,benneke2019sub}. Atmospheric retrieval is the act of obtaining the physical and chemical characteristics of an atmosphere based on observations. Observations of exoplanet atmospheres often have a poor signal to noise and different combination of atmospheric parameters can often be compatible with the data. In order to quantify these degeneracies, atmospheric retrievals are often carried out by computing thousands of models allowing to explore a large parameter space and quantify the degeneracies between parameters. However, when performing retrievals on a data set, one has to first carefully choose the forward model that will be used for the data analysis. Such a choice includes a series of assumptions allowing one to reduce the parameter space to explore. These assumptions can sometimes bias the retrieval results towards unrealistic values, leading to precise, but wrong inferences about the atmospheric composition of an exoplanets atmosphere. 

In this paper we will look specifically at the assumption that atmospheres are horizontally homogeneous. Such a 1-D assumption has been widely used to study exoplanet atmospheres \citep[e.g.][]{kreidberg2014precise,barstow2016consistent,wakeford2017complete,evans2017ultrahot,evans2018optical}. However, it is clear that the atmosphere of exoplanets are often intrinsically three-dimensional. In the specific case of tidally locked hot Jupiters, the large irradiation received on the dayside and the poor day-to-night heat redistribution can lead to variations of up to a thousand degrees between the dayside and the nightside \citep{showman2002atmospheric,knutson2007map,parmentier2017exoplanet}. 

Recently, \citet{feng2016impact} showed that when retrieving the spectrum of an inhomogeneous planet with a 1-D model the inferred molecular abundances could be strongly biased. In several cases, this could even lead to spurious molecular detections. They concluded that such biases can already be seen in current data, but will likely be much stronger when better observations become available. For their JWST-like simulations ($\lambda$ = 1 -- 10 $\mu$m) they find that 2-D effects become significant if the temperature contrast between two thermal profiles observed at planetary quadrature is larger than 40\%.

This study follows-up on the one of \citet{feng2016impact}. Our goal is understand the mechanisms leading 1-D retrieval models to infer the presence of spurious atmospheric constituents when applied to observations from inhomogeneous atmospheres. For this we first develop a simple analytical model of the effect that an inhomogeneous thermal structure can have on the emission spectrum of a planet (Section \ref{section2}). This allows us to quantify the signal-to-noise ratio necessary to be able to measure the presence of inhomogeneities on the planetary disk. We then look more specifically than \citet{feng2016impact} on the consequences of an inhomogeneous atmosphere on the retrieved abundances when different instrument modes are used, focusing particularly on JWST/NIRSpec (Sections \ref{section3} and \ref{section4}). We find that that in most cases, retrieval models will infer biased molecular abundances from the JWST/NIRSpec observations of hot Jupiter atmospheres unless the inhomogeneous nature of the atmosphere is taken into account. Finally, in Section~\ref{sec:Bias} we uncover the mechanism responsible for the spurious inferences of molecular constituents in planetary atmospheres. We conclude that considering the presence of an inhomogeneous atmosphere will be crucial to avoid spurious molecular detections and obtain accurate molecular abundances in the JWST era.

\section{Detectability of TP inhomogeneities}
\label{section2}

\begin{figure}
    \includegraphics[scale=0.45]{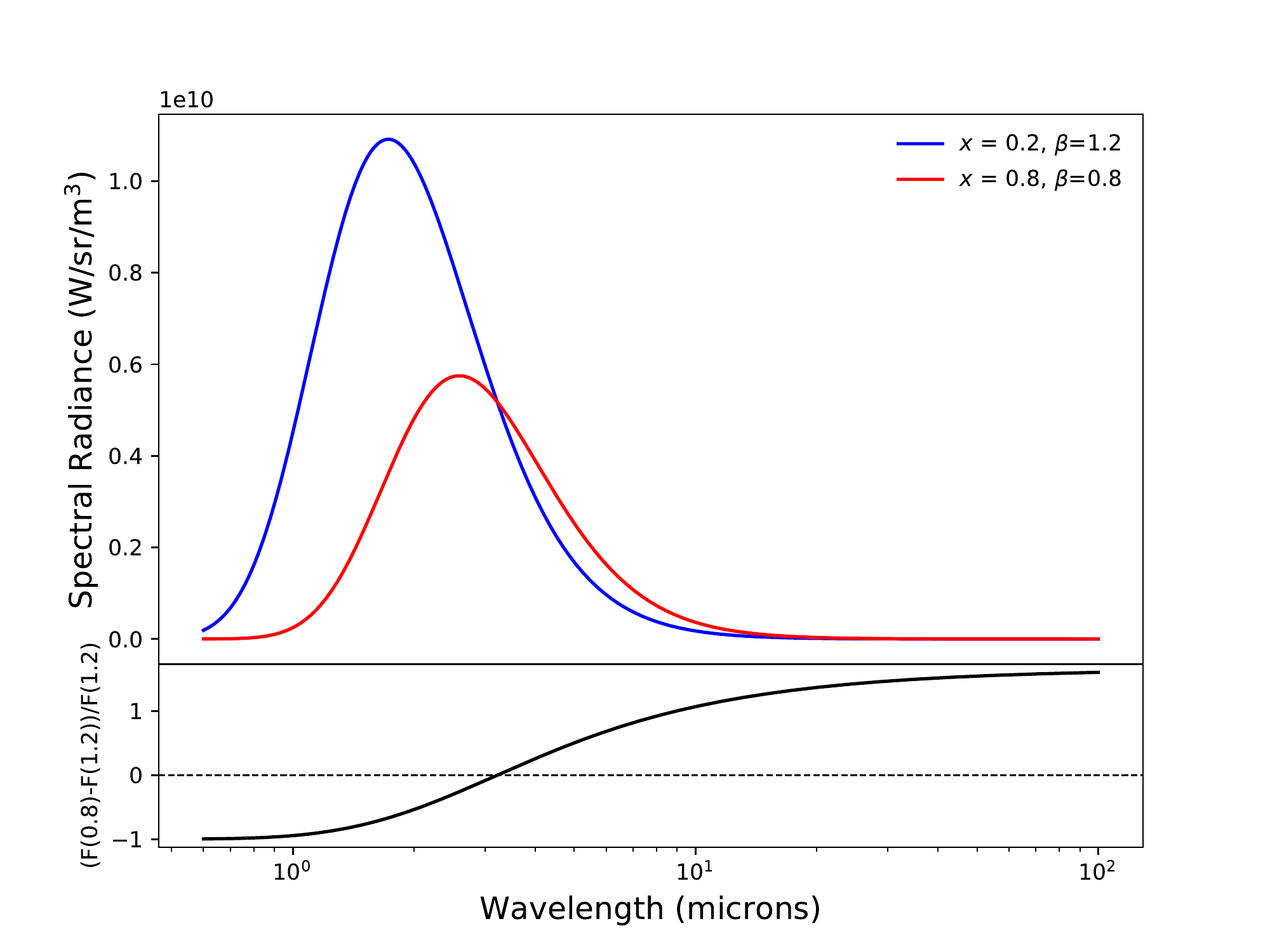}
    \caption{Plot showing two planetary spectra considering that a part $x$ of the atmosphere emits like a blackbody and the other part is not emitting any light. The blue curve has a $\beta$ value of 1.2 and the red curve has the value of 0.8, the function has been plotted using Eq \ref{flux_soln}. The hotter profile covers a smaller fraction of the surface ($x = 0.2$), the opposite is true for the colder profile ($x = 0.8$). For a given wavelength, there exists more than 1 combination of these parameters which give the same flux output. Note: that at a given wavelength, there exists multiple atmospheric scenarios which can give the same flux output.}
    \label{fig:bb_cross}
\end{figure}

We start by deriving an analytical model to determine what precision and wavelength range are necessary to detect the presence of horizontal inhomogeneities based on a single spectrum only. In order to simplify the problem we will assume that a fraction $x=x_0$ of the visible disk emits like a blackbody at the temperature $T_{\rm hot}=T_{\rm 0}$ whereas the rest of the disk is not emitting any light (i.e. $T_{\rm cold}$=0). The resulting spectrum, plotted in Fig. \ref{fig:bb_cross} is therefore a blackbody spectrum diluted by a factor $x_0$.

As shown by the lines crossing in Fig. \ref{fig:bb_cross}, for a given wavelength $\lambda_0$, the flux $F(\lambda_0)$ that is observed could also correspond to a planet with a hotter hot part covering a smaller area or a colder hot part covering a larger area. Indeed if observations at a single wavelength are made there is an intrinsic degeneracy between the temperature of the hot region and the area covered by it. If we write the hot temperature as $T_{\rm hot}=\beta T_{0}$, we can write that the flux emitted by the planet is: 
\begin{equation}
    F(\lambda) = xB(\beta,\lambda) \times \frac{\pi a^2}{D^2},
\end{equation}
Where $a$ is the planet radius and $D$ is the distance from the Earth and $B(\beta, \lambda)$. For convenience we will now assume $\frac{\pi a^2}{D^2}=1$. $B(\beta,\lambda)$ is the Planck function taken at a temperature $T=\beta T_0$: 
\begin{equation}
    B(\beta,\lambda) = \frac{2hc^2}{\lambda^5}\frac{1}{e^{\frac{hc}{\lambda k\beta T_0}}-1},
\end{equation}
The total flux emitted by the planet at a given wavelength can be written:  
\begin{equation}
    F(\lambda) = x_0B(\beta_0,\lambda),
\end{equation}
For a specific wavelength $\lambda_0$, the ensemble of models that can match the planetary flux is described by :
\begin{equation}
    F(\lambda_0) = x_0B(\beta_0,\lambda_0) = xB(\beta,\lambda_0).
\end{equation}
This can also be written as a relationship between $\beta$ and $x$:
\begin{equation}
\label{x_relation}
    x = \frac{x_0B(\beta_0,\lambda_0)}{B(\beta,\lambda_0)}.
\end{equation}

We can see that at a given wavelength $\lambda_0$, there exists a continuous ensemble of solutions ($\beta$,$x$) that can match the planetary flux. However, as is seen if Fig. \ref{fig:bb_cross}, the different solutions will diverge and become distinguishable away from $\lambda_0$. We now wonder what additional observations could allow us to distinguish between the real planet parameters $x_0, \beta_0$ and the ensemble of solutions degenerate at $\lambda=\lambda_0$. We first write the flux of one of these solution as:

\begin{equation}
\label{flux_soln}
    F(\lambda) = xB(\beta,\lambda) = \frac{x_0B(\beta_0,\lambda_0)}{B(\beta,\lambda_0)} \times B(\beta,\lambda)\, .
\end{equation}
Now let us imagine that we have two observations, one at a wavelength $\lambda_0$ and one at a wavelength $\lambda_0+\Delta \lambda$. We want to derive a criteria to understand what signal-to-ratio would be necessary to disentangle between the true solution and a planet model where the temperature of the hot part of the atmosphere differ from the cooler part by $\Delta \beta$. 

If the difference between the profiles is larger than the error on the measurement ($\sigma_{F_{\rm p}}$) then the model at $\beta_0+\Delta\beta$ will not be a good fit to the observations. This can be written as: 
\begin{equation}
\label{eq_7}
|F(\lambda_0+\Delta \lambda, \beta_0, x_0)-F(\lambda_0+\Delta \lambda, \beta_0+\Delta\beta, x_0)|>\sigma_{F_{\rm p}}
\end{equation}

We now suppose that we try to disentangle two models that differ by a small value $\Delta\beta$. We further assume that the model fluxes are approximately linear over the wavelength range $\Delta \lambda$ considered. This last approximation would hold best away from the peak of the Planck function. For a $\Delta\lambda$ of $1\mu m$ the error induced by the linearisation is going to be at maximum of a factor 2 to 3. As seen later, even a factor 2 is a good enough as the flux variations can span orders of magnitude difference when varying $\beta$ and $\Delta\lambda$. 

We perform a Taylor expansion on Eq \ref{eq_7} of order one (see Eq \ref{dqdw} and Eq \ref{dqdwdb}) as a function of $\beta$ and $\lambda$, which leads to the following equation:
\begin{equation}
\label{eq_8}
       \frac{\partial}{\partial\beta}\Big(\frac{\partial F}{\partial\lambda}\Big)_{\lambda_0,\beta_0}\Delta\lambda\Delta\beta > \sigma_{F_{p}} \text{.}
\end{equation}
We now divide both sides of \ref{eq_8} by the flux ($F_p$) and define the signal-to-noise ratio as $SNR=F_p/\sigma_{\rm F_p}$, hence we obtain a criterion for the detectability of spatial inhomogeneities in a planet atmosphere:
\begin{equation}
\label{criterion}
    \frac{1}{F_{p}}\frac{\partial}{\partial\beta}\Big(\frac{\partial F}{\partial\lambda}\Big)_{\lambda_0,\beta_0}\Delta\lambda\Delta\beta > \frac{\sigma_{F_{p}}}{F_{p}}\approx\frac{1}{SNR}.
\end{equation}
We note that this equation can be seen as a function of $\lambda_0$ and $\beta_0$ or as a function of $\lambda_0$ and $x_0$ since $\beta_0$ and $x_0$ are related by Eq \ref{x_relation}.

\begin{figure}
    \centering
    \includegraphics[scale=0.35]{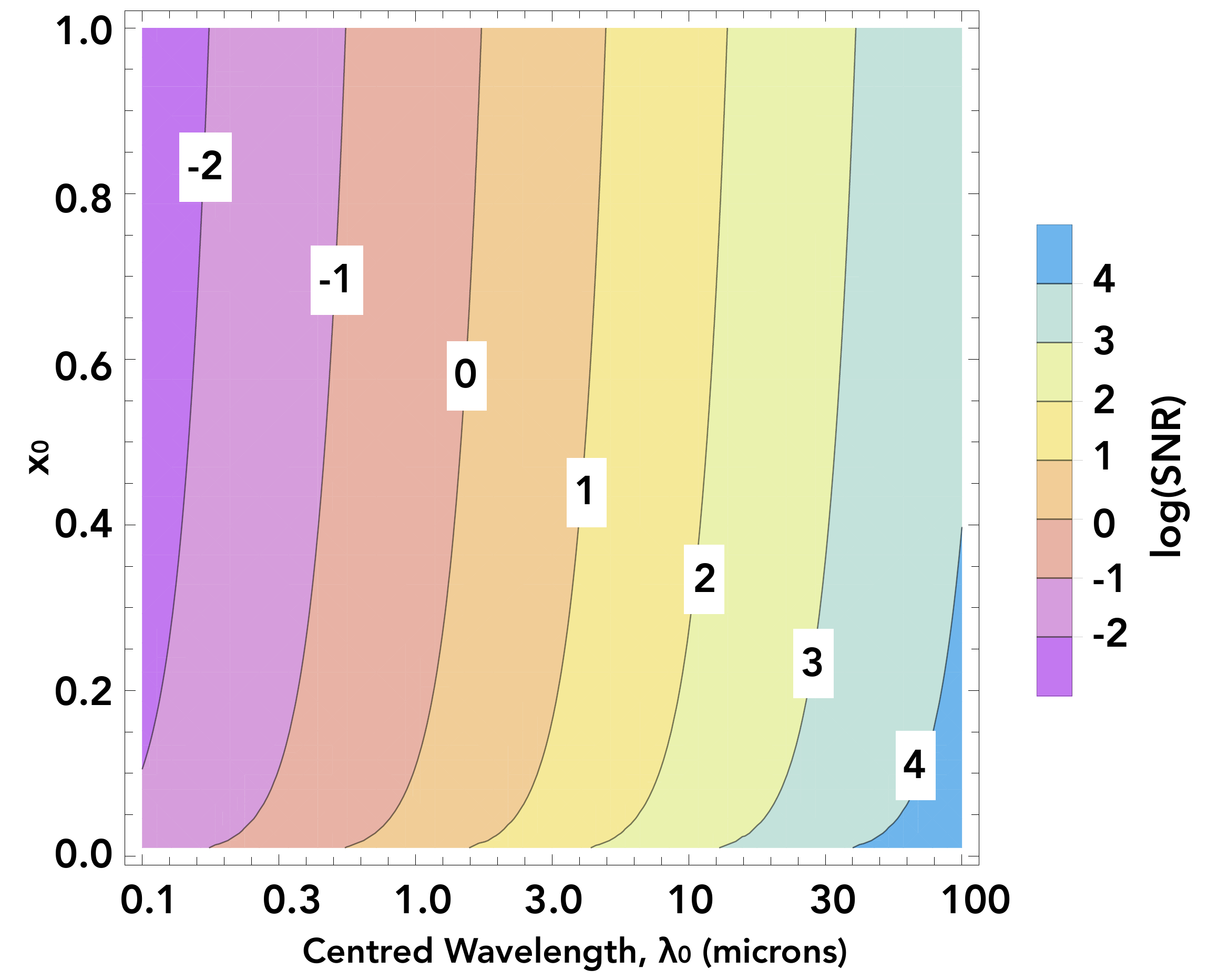}
    \caption{
    Ability to disentangle between a planetary spectrum with a hot region with a temperature of 1400K covering a fraction $x_0$ of the surface and a planetary spectrum with a colder region of $\beta_0$ + $\Delta \beta$ = 0.8, where $\beta_0$ = 1. Hence the colder region has a temperature of ($\beta_0$ + $\Delta \beta)\times 1400 = 1120$K. For the colder planetary case the emitting part of the planet covers a larger fraction in order to have the same flux at $\lambda_0$. 
    The SNR are calculated following Eq. \ref{criterion} assuming a wavelength range $\Delta\lambda=2\mu m$ centered at different $\lambda_0$ as shown by the x-axis. The plotted threshold can be extrapolated to different contrasts by noting that the SNR scales inversely with the temperature contrast $\Delta \beta$. Note that shorter wavelengths are more important in being able to detect 2-D effects. Increasing the contrast of the profiles shifts the contour lines to longer wavelengths, however the effect is still marginal.}
    \label{fig:snr}
\end{figure}

We use this criterion to show the ability to disentangle 2-D structures for a planet with a hot isotherm of 1400K covering different fraction of the visible planetary disk, $x_0$ in Fig. \ref{fig:snr}, we chose to model a wavelength coverage of $\Delta\lambda=2\mu m$ to represent the majority of bandpasses covered by JWST instruments. We see from the plot that the contour lines do not vary much with the hot profile covering a fraction $x_0$. Secondly we see that as we increase the central wavelength, the ability to disentangle the inhomogeneous structures becomes harder, requiring a higher SNR. Lastly, the contour lines scale with the temperature contrast $\Delta \beta$ as seen from Eq \ref{criterion}, with a larger temperature contrast shifting the contour lines to the right, hence requiring a lower SNR, this effect is marginal. 
 

We can see that the contour line at 1 micron corresponds to a SNR $\sim$ 1 for a low contrast ($\Delta \beta$ = 0.2) This means that with the observation centred at 1 micron, and a wavelength coverage of 2 microns ($\Delta \lambda$), we are able to easily disentangle 2-D effects even if the secondary eclipse is measured with an SNR of 1 only. The SNR required increases as we move to longer wavelengths, where at 10 microns we need a SNR in the range 10 -- 100. To conclude, we find that observing shorter wavelengths and a large wavelength coverage is best to disentangle temperature inhomogeneities from a disk-averaged spectrum.

\section{JWST Simulations}
\label{section3}
\subsection{Forward Model}
We use the \textbf{N}on-linear optimal \textbf{E}stimator for \textbf{M}ultivariat\textbf{E} spectral analy\textbf{SIS} code \textsc{NEMESIS} \citep{irwin2008nemesis} to compute our model spectra. \textsc{NEMESIS} uses the correlated-$k$ approach \citep{lacis1991description} to model the spectra. We built our $k$-distribution look up tables to have a resolution (FWHM) of $\Delta\lambda = 0.02 \mu m$, using line data from various sources. For this study the molecules used are: H$_2$O \citep{barber2006high}, CO \citep{rothman2010hitemp}, CO$_2$ \citep{tashkun2011cdsd} and CH$_4$ \citep{yurchenko2014exomol}.  As we are modelling an atmosphere which is H$_2$-dominated we also modelled the H$_2$-H$_2$ and H$_2$-He collisionally-induced absorptions \citep{richard2012new}. 

\subsection{TP profile parameterisation}
\label{tp_param}
To model our thermal profiles we employ the five-parameter, three-channel description of the incoming and outgoing fluxes presented in \citet{parmentier2014non}. 
Briefly, the temperature as a function of the thermal optical depth, $\tau$, is given by

\begin{equation}
\label{tp_profile}
    T^4(\tau) = \frac{3T^{4}_{\text{int}}}{4}\Big(\frac{2}{3}+\tau\Big) + \frac{3T^{4}_{\text{irr}}}{4}(1-\alpha)\zeta_{\gamma_1}(\tau) + \frac{3T^{4}_{\text{irr}}}{4}\alpha\zeta_{\gamma_2}(\tau),
\end{equation}
with $\zeta_{\gamma_i}$ given by

\begin{equation}
    \zeta_{\gamma_i} = \frac{2}{3} + \frac{2}{3\gamma_i}[1 + (\frac{\gamma_i \tau}{2} -1)e^{(-\gamma_i \tau)}] + \frac{2\gamma_i}{3}(1 - \frac{\tau^2}{2})E_2(\gamma_i \tau).
\end{equation}
The parameters $\gamma_{1}$ and $\gamma_{2}$ are the ratios of the mean opacities in the two visible streams to the thermal stream: $\gamma_{1} = \kappa_{v1} / \kappa_{IR}$ and $\gamma_{2} = \kappa_{v2} / \kappa_{IR}$. The parameter $\alpha$ has a range between 0 and 1 and is the weighting used between the two visible streams, $\kappa_{v1}$ and $\kappa_{v2}$. $E_2(\gamma_i \tau)$ is the second order
exponential integral function. $T_{\text{int}}$ is the internal temperature of the planet and it is usually a fixed value as it does not have much effect on the spectra as long $T_{\text{int}}^4 \ll T_{\text{irr}}^4$. Here we have fixed it to a value of 200 K following \citet{line2013systematic}. $T_{\text{irr}}$ is the radiative equilibrium temperature and is of the form

\begin{equation} \label{betadefine}
    T_{\text{irr}} = \beta\left(\frac{R_*}{2a}\right)^{1/2}T_*,
\end{equation}

where $R_*$ and $T_*$ are the stellar radius and temperature respectively, $a$ is the semi-major axis of the planets orbit, and $\beta$ is a parameter which is used as a proxy for determining how hot or cold a profile is. Finally the parameter $\tau$ is the grey infrared opacity 
\begin{equation}
    \tau = \frac{\kappa_{IR}P}{g},
\end{equation}
where $\kappa_{IR}$ is the mean infrared opacity, $P$ is the pressure in the atmosphere and $g$ is the surface gravity, which we consider to be constant with a value of $\log$(g) = 3.69 cm s$^{-2}$ at the 1 bar level (calculated using the mass and radius presented in Table \ref{tab:model_param}).  The profile therefore has five free parameters: $\alpha$, $\beta$, $\kappa_{IR}$, $\gamma_{1}$ and $\gamma_{2}$.

\begin{figure}
    \centering 
    \includegraphics[scale=0.55]{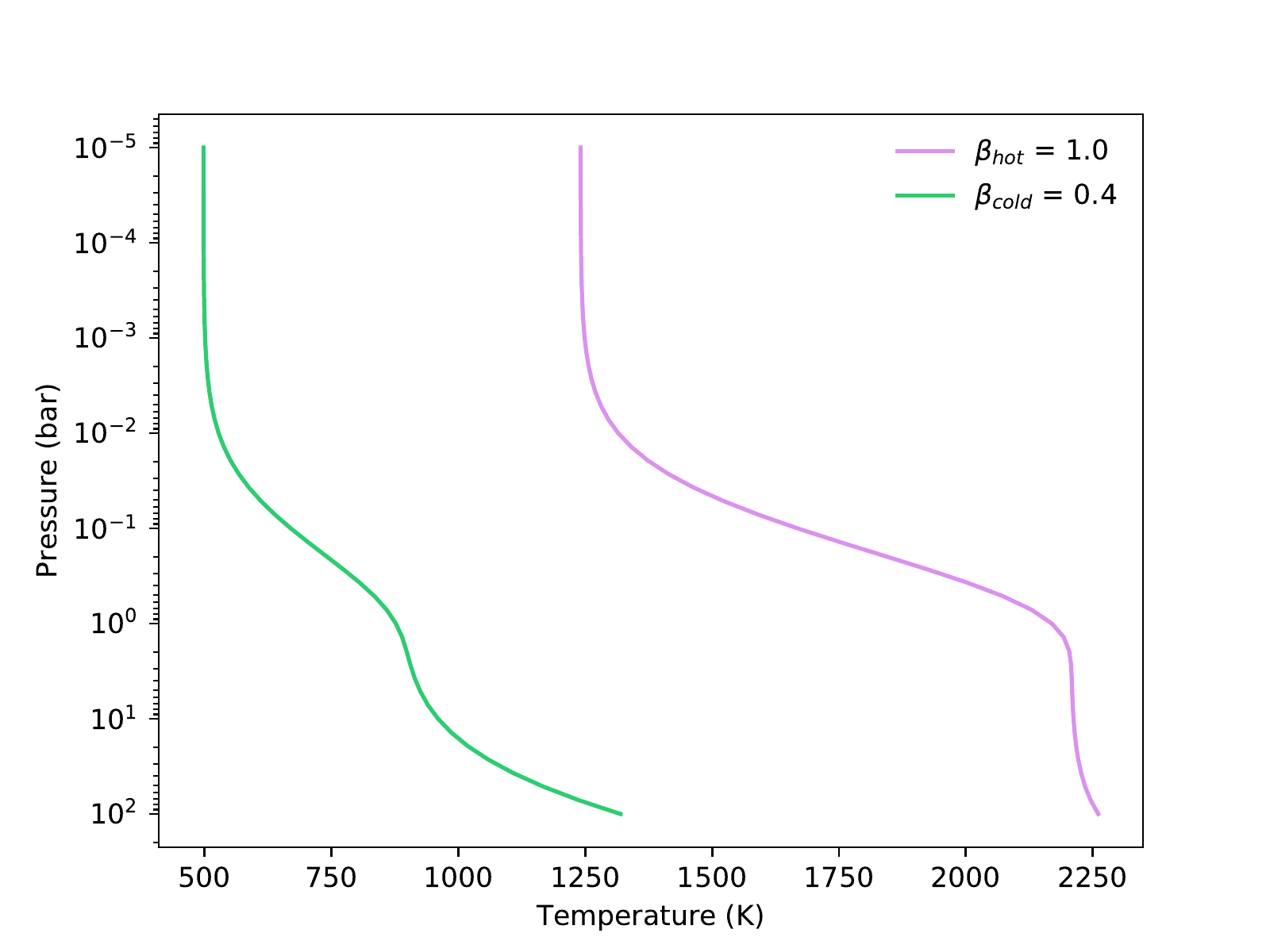}
    \caption{The temperature-pressure (TP) profiles used to generate the model atmosphere for this study. The parameterisation of the TP profiles are presented in \citet{line2013systematic} and \citet{parmentier2014non}. We present a brief description in Section~\ref{tp_param}. }
    \label{fig:tp_example}
\end{figure}

\subsection{Composite Spectra}
\begin{figure}
    \centering 
    \includegraphics[scale=0.55]{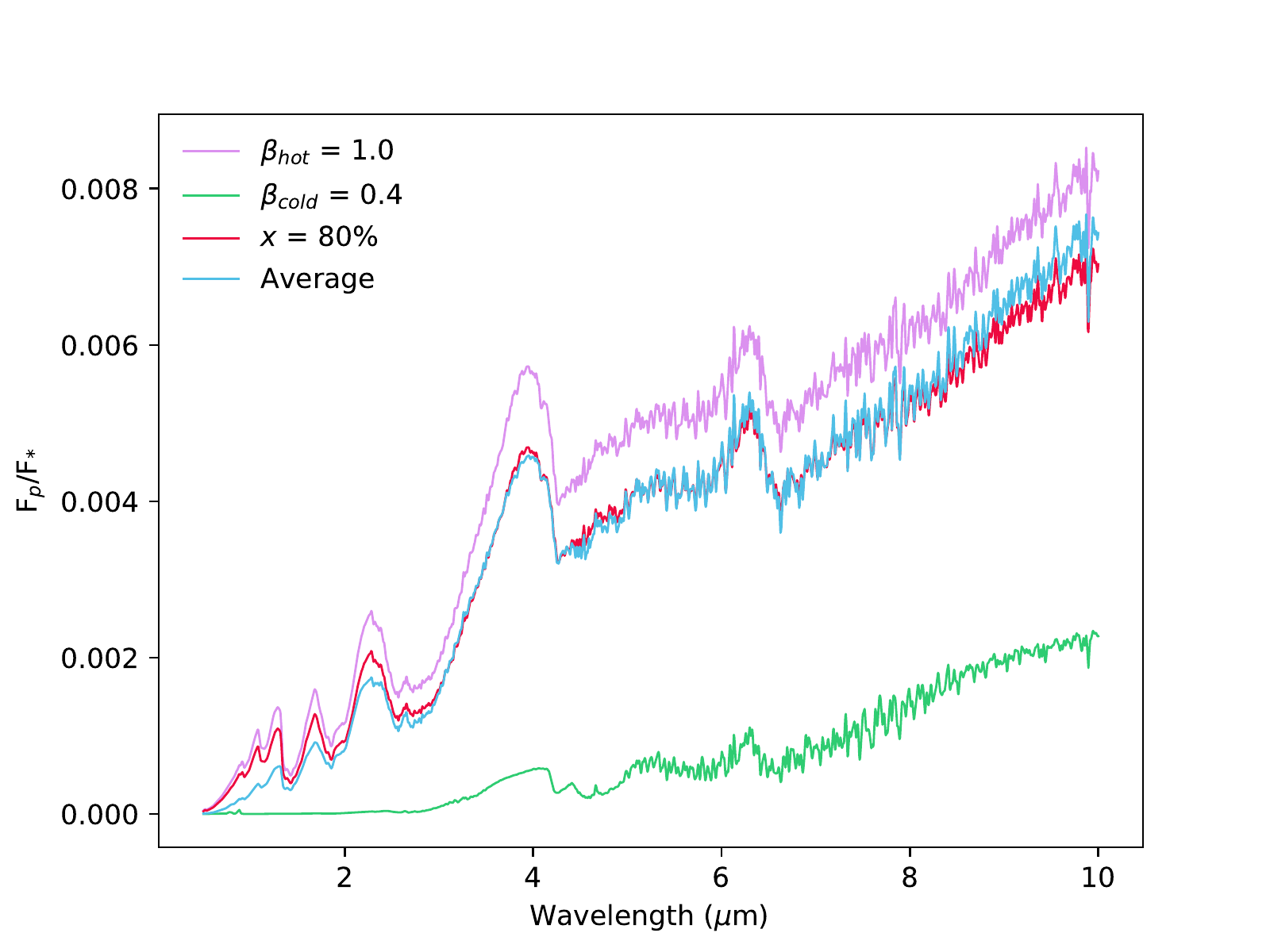}
    \caption{Spectra for the hot ($\beta_{\text{hot}}=1$, magenta) and cold ($\beta_{\text{cold}}=0.4$, green) profiles, alongside the composite spectrum (red) produced with a weighting of $x$=
    80\%. This composite spectrum was binned and noise added to produce synthetic data used in the retrieval study. For comparison, we also show (in blue) a spectrum generated using the average of the hot and cold TP profiles ($T_{\text{average}}^4 = \frac{1}{2}(T_{\text{hot}}^4 + T_{\text{cold}}^4$)). Note the small, but visible differences between the composite and average spectra, especially at shorter wavelengths. }
    \label{fig:spec_example}
\end{figure}

Similarly to \citet{feng2016impact} we build a uniformly mixed, cloud-free model atmosphere assuming that part of the atmosphere is hotter than the rest. We now want to model the spectrum of a horizontally inhomogeneous planet atmosphere. For this we assume that a projected fraction $x$ of the planet hemisphere we are looking at is hotter. To do this we employ a linear weighting scheme
\begin{equation}
    F_{\text{weighted}} = xF_{\text{hot}} + (1-x)F_{\text{cold}} \text{,}
    \label{eq:eq_1}
\end{equation}
where $F_{\text{hot}}$ is the spectrum generated from the hot 1-D TP profile, $F_{\text{cold}}$ is the spectrum generated from the cooler 1-D TP profile and $x$ represents the weight applied to the hotter profile and is a free parameter in our retrievals. We vary the weights used to generate the observed spectra to be $x$ = 80\%, 60\%, 40\% and 20\%. Figure \ref{fig:spec_example} gives an example of how the observed spectrum was generated for a weighting of 0.8. In physical terms, if the hot and cold profiles correspond to the day- and nightside respectively, then the weights $x$ = 80\%, 60\%, 40\% and 20\% represent approximately a planet observed at phase angles of $\theta = 144^{\circ}$, $108^{\circ}$, $74^{\circ}$ and $36^{\circ}$ respectively (also $360^{\circ}-\theta$).  

\subsection{Treatment of angles}
We appreciate that when observing a planet at different points in its phase there is a need to consider different zenith angles at which a hot spot might be observed. However, we chose to fix the zenith angle to be 45$^{\circ}$ for all simulations to reduce the layer of complexity. It has been shown by \citet{irwin25d} that for a 1-D retrieval, the assumed zenith angle has second order effects on the temperature-pressure profile; as our technique is a combination of two 1-D models, it should not have an extreme influence. 

As we are focusing on the abundances retrieved rather than the temperatures, having a fixed zenith angle is a useful approximation. If the hotter region of the planet was modelled with a varying viewing angle, the retrieval would have to adjust the TP profile shape to compensate, which might effect the retrieved abundances.

\subsection{WASP-43b}
The planet and stellar parameters are based on the exoplanet WASP-43b and its star discovered by \citep{hellier2011wasp}, we present the key parameters used in the modelling in Table~\ref{tab:model_param}. We use these parameters to generate a stellar spectrum from the Kurucz \textsc{ATLAS9} model atmospheres \citep{castelli2004new}. We use the average TP profile and assume equilibrium chemistry to calculate the volume mixing ratios (VMR) using \textsc{GGchem} \citep{woitke2018equilibrium} and take the approximate values at 100 mbars to create a uniformly mixed atmosphere; these values are also listed in Table~\ref{tab:model_param}. We appreciate that the atmospheres may not be uniformly mixed and that the different profiles may have different chemistry if the atmosphere is not quenched \citep{changeat2019towards}. We aim to explore this in a later study. 
For this study we use two TP profiles with a large contrast, representative of the expected day-night contrast for some hot Jupiters. Specifically, our model atmosphere is based on WASP-43b, which is known to have a day-night contrast of $\sim$1000K \citep[][]{kataria2015atmospheric,mendoncca2018revisiting,morello2019independent}. The cold and hot profiles are consistent with retrieved TP profiles of the nightside and dayside respectively found in \citet{stevenson2017spitzer}. Other hot Jupiters have been shown to have even larger day-night contrasts \citep{komacek2016atmospheric,parmentier2017exoplanet}, so our present study is by no means an extreme case.

\begin{table*}
   \centering
\begin{tabular}{ll|ll|ll|ll}
    \hline
    \multicolumn{2}{c}{Chemistry}      & \multicolumn{2}{c}{Planet} &   \multicolumn{2}{c}{Star} &  \multicolumn{2}{c}{TP}  \\
    \hline
    $\log$H$_2$O     & -3.42 & $R_{\text{P}}$ $(R_{\text{J}})$ & 1.036 & $R_*$ $(R_\odot)$ & 0.667& $\log\kappa_{\text{IR}}$ & -1.0\\
    $\log$CO & -3.34 & $M_{\text{P}} (M_{\text{J}})$ & 2.052 & $T_*$ (K) & 4520 & $\log\gamma_{1}$ & -1.0\\
    $\log$CO$_2$ & -6.63 & $\log$(g) (cm s$^{-2}$)& 3.69  & $M_* (M_\odot)$& 0.717 & $\log\gamma_{2}$ & -1.0\\
    $\log$CH$_4$ & -8.20 & $T_{\text{int}}$ (K) & 200 &[Fe/H] & -0.01& $\alpha$ & 0.5\\
     & & $a$ (AU) & 0.01526 &&& $\beta_{\text{hot}}$ & 1.0\\
     & & & &&& $\beta_{\text{cold}}$ & 0.4
\end{tabular}
\caption{Values used to create the forward model atmospheres. In column 1 we present the volume mixing ratios of the molecules which make up the atmosphere, with the rest of the atmosphere consisting of 85\% H$_2$ and 15\% He. In column 2 and 3 we present the physical parameters of the planet WASP-43b and the star WASP-43 respectively \citep{gillon2012trappist}. In column 4 we present the parameters used to create the two temperature-pressure profiles used in this study.}
    \label{tab:model_param}
\end{table*}

\subsection{Application to NIRSpec}
Once the weighted spectrum is generated (i.e the red spectrum shown in Fig.~\ref{fig:spec_example}) it is binned to the wavelength grid of the different instruments in this study. We used \textsc{PandExo} \citep{batalha2017pandexo} to generate the wavelength grid and noise for the NIRSpec Prism (0.6 -- 5.3 $\mu$m), NIRSpec G395M (2.87 -- 5.10 $\mu$m), NIRSpec G235M (1.66 -- 3.07 $\mu$m) and NIRSpec G140M (0.97 -- 1.84 $\mu$m), this is done for the integration time of the secondary eclipse of WASP-43b. For the NIRSpec Prism (nominal resolving power $\sim$100) we use the resolution of the largest bin step as shown in \citet{krissansen2018detectability} and for the different NIRSpec gratings (nominal resolving power $\sim$1000) we use a resolution of R=100. 

We chose to focus this study on the various NIRSpec instrument modes (G140M, G235M, G395M and Prism) as these will be used for transiting exoplanet observations and also provide a large overlapping wavelength grid. We decided to also use the medium resolution gratings (R $\sim$ 1000) as opposed to the higher resolution gratings (R $\sim$ 2700) to test the lower limits of the resolution obtained by JWST. We ask, is the wavelength coverage of these instruments good enough to break the degeneracy, hence requiring a 2-D approach to better analyse the observations? And, if it is possible to retrieve the 2-D effects, is there evidence to support a 2-D approach over a 1-D approach?

We note that the NIRSpec Prism will saturate for bright targets and so for brighter objects an observer would opt to use the gratings \citep{stevenson2016transiting}. The gratings provide a much smaller wavelength coverage so an observer may want to combine observations of different gratings. We investigate if by combining G235M and G395M we can recover information better than if we just had observations of the individual gratings; in addition, we test if we are biasing our retrievals when we do not combine the instruments. It has been shown that combining observations can introduce biases in the retrieval and so systematic correction also needs to be considered \citep{barstow2015transit}.

\begin{figure}
    \centering
    \includegraphics[scale=0.25]{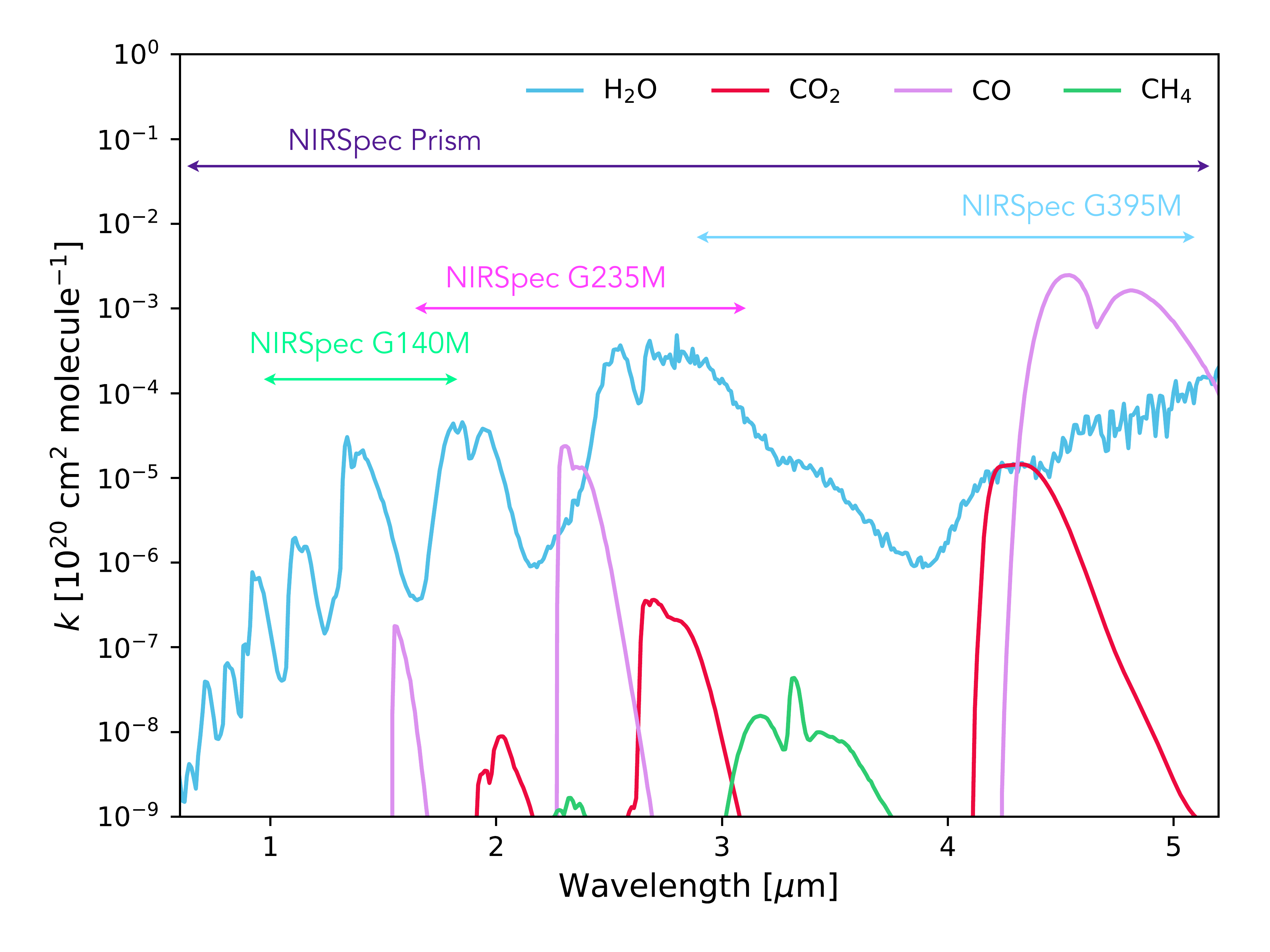}
    \caption{The weighted-by-abundance absorption coefficients of the molecules used in this study. We have highlighted the wavelength coverage of each of the instruments. The absorption coefficients are calculated at a temperature of 1450K and pressure of 0.1bar.}
    \label{fig:molecules}
\end{figure}

\subsection{Retrieval}
For the retrievals conducted in this study we have wrapped the forward-modelling component of \textsc{NEMESIS} in a Bayesian framework, namely using a nested sampling approach \citep{feroz2008multimodal,feroz2013importance}, which we implemented using \textsc{pymultinest} \citep{buchner2014x}. We use a nested sampling approach due to its ability to produce posterior distributions of the parameters and also for its efficiency in calculating the Bayesian evidence, which can be used in model comparison and selection.

We retrieve on the model atmospheres with 3 different retrieval styles: a 1-D model (i.e. $F_{\text{1-D}} = F_{\text{average}}$, where $F_{\text{average}}$ is the disc averaged spectrum assuming a homogeneous temperature profile), a 1-D model with a free parameter, which is able to dilute the spectrum ($F_{\text{dilution}} = sF_{\text{average}}$) and then a model with 2 TP profiles which follows Eq \ref{eq:eq_1}, this model is 2-D. These will be referred to as 1-D, dilution and 2-D styles respectively from now on.

\subsection{The Dilution Retrieval}
Here we employ a method we dub the "dilution retrieval" where we consider the flux coming from the hot profile alone, effectively considering the colder profile to exhibit zero flux output. Hence it follows the style of Eq~\ref{eq:eq_1} with the second term equalling zero i.e., 
\begin{equation}
    F_{\text{dilution}} = sF_{\text{hot}} + (0 \times F_{\text{cold}}) \text{,}
\end{equation}

The parameter $s$ effectively scales the 1-D forward model $F_{\text{hot}}$ to fit to the observations. The forward model will be of a higher temperature than the disc average as it is seeing a smaller emitting area on the disc of the planet, we show this in Section \ref{section2}. The scaling is wavelength independent and mutes the features equally.

The "dilution" factor, $s$, retrieved is consistent with the weighting of the hot profile. The technique gives freedom for the retrieval to adjust to accommodate systematics in the observations as well as account for the non 1-D nature of the spectrum.
This technique is useful to use as it reduces the number of free parameters needed to conduct a meaningful retrieval simulation (extra parameter compared to the parameters needed to simulate a second TP profile). This makes sense to use when considering an emission spectrum where the colder part of the planet is much colder than the hotter part, hence being completely dominated by the spectral features in the hotter region. For the large temperature contrast, the colder region could also be seen as a cloudy profile \citep{irwin20192pt5D}. 

\subsection{Bayesian Evidence and Model Selection}

We use Bayesian model comparison to evaluate the statistical significance of one retrieval model over another \citep[e.g.][]{Ford2007,schulze2012bayesian}.
We compute the Bayes factor $\ln B = \ln Z_2 - \ln Z_1$, where $Z_1$ and $Z_2$ are the evidences for the 1-D and 2-D models, respectively, as calculated by the nested sampling algorithm.
If the Bayes factor is significantly less than unity ($\ln B<0$), the data supports model 1 (1-D model). If it is significantly greater than unity, the reverse is true. If $|\ln B|$ is close to zero, one cannot discriminate between the two models. It is common practice, when using the Bayes factor for model comparison, to interpret the results using the adapted Jeffrey scale \citep{kass1995bayes}. For example, the preference for model 2 over model 1 is considered statistically significant only if $\ln B > 3$. However, our target audience of astronomers and planetary scientists is most familiar with degrees of significance expressed in terms of the number of standard deviations, $\sigma$ (these are directly related to confidence intervals if the distribution in question is Gaussian). To make our results easily accessible to this audience, we converted our $\ln B$ values to a number of standard deviations, following the prescription of  \cite{trotta2008bayes}. We then consider a particular model to be strongly favoured if it is supported at the $>3.6\,\sigma$ level.

\section{Results}
\label{section4}
We aim to show which atmospheric retrieval style was most appropriate in retrieving on an atmosphere which has 2 TP profiles. The results of these simulations are shown in Table \ref{tab:80_results} and are for the different planet types: with 80\%, 60\%, 40\% and 20\% weighting applied to the hot profile, where the hot profile has a $\beta_{\text{hot}}$ = 1.0 and the cold profile has a $\beta_{\text{cold}}$ = 0.4. These results are instructive: they show when there is evidence to suggest that 2 TP profiles and also a more simple dilution retrieval will be needed to perform a more robust analysis of data obtained from different NIRSpec observing modes. Alongside the $\sigma$-significance of the retrieval method, we have presented whether or not the retrieved chemistry is biased (B) or unbiased (U). For this study, we define an unbiased retrieval to be one where the volume mixing ratio (VMR) of the species that have features in the wavelength region observed were constrained to values within 2-$\sigma$ of the input values. We define a biased retrieval to be one where one of the molecular abundance is constrained outside 2-$\sigma$ of the true value. 

In the next subsections we describe our results for different instruments and different planet models. For each observing mode apart from Prism and G140M, we have presented the best fitting retrieved models and accompanying posterior distributions for the 60\% case in the Appendix. For Prism we show the best fitting retrieved models and accompanying posterior distributions for the 20\% and 80\% case in the main text, we also show the G140M 60\% case in the main text. The rest of the results are just described.

\begin{table*}
\begin{center}
\begin{tabular}{cccccccccccc}
&  &$x$ = 80\%& & & & &&$x$ = 60\%&  & \\
\hline
Observing Mode & \multicolumn{2}{c}{2-D} & \multicolumn{2}{c}{Dilution} & 1-D  &Observing Mode   & \multicolumn{2}{c}{2-D}& \multicolumn{2}{c}{Dilution}& 1-D \\
\hline
NIRSpec G140M & 1.85$\sigma$&  {U} & 2.11$\sigma$ & {U} &   {B} & NIRSpec G140M   & 4.58$\sigma$ & {U} & 4.52$\sigma$ & {U} &   {B} \\     
NIRSpec G235M & 1.97$\sigma$ &  {U} & $<$ 1$\sigma$  &   {U} &  {U} & NIRSpec G235M  & 4.56$\sigma$&  {U} & 4.46$\sigma$ & {U} &   {B}  \\      
NIRSpec G395M & $<$ 1$\sigma$&  {U} & $<$ 1$\sigma$ & {U} &   {U} & NIRSpec G395M &  $<$ 1$\sigma$ & {U} & $<$ 1$\sigma$ & {U} &   {B}  \\ 
NIRSpec Prism & 9.89$\sigma$ & {U} &9.64$\sigma$ & {U} &  {B} & NIRSpec Prism &   17.06$\sigma$ & {U}  & 17.00$\sigma$ & {U} &  {B}  \\
G235M + G395M  & 11.11$\sigma$ & {U} &11.12$\sigma$ & {U} &  {B} & G235M + G395M &   18.94$\sigma$ & {U}  & 18.86$\sigma$ & {U} &  {B}  \\
\hline
&  &$x$ = 40\%& & & & &&$x$ = 20\%&  & \\
\hline
Observing Mode  & \multicolumn{2}{c}{2-D}& \multicolumn{2}{c}{Dilution}& 1-D  & Observing Mode  & \multicolumn{2}{c}{2-D}& \multicolumn{2}{c}{Dilution} & 1-D \\
\hline
NIRSpec G140M & 4.12$\sigma$ & {U} & 4.18$\sigma$&  {U} &   {B} & NIRSpec G140M &  4.53$\sigma$ & {U} & 4.61$\sigma$ & {U} &   {B} \\
NIRSpec G235M & 4.42$\sigma$ & {U} & 4.32$\sigma$ & {U} &   {B} & NIRSpec G235M &   5.18$\sigma$ & {B} & 5.23$\sigma$ & {B} &   {B}  \\ 
NIRSpec G395M & $<$ 1$\sigma$&  {B} & $<$ 1$\sigma$ & {B} &   {B} & NIRSpec G395M &  2.77$\sigma$&  {U} & 2.36$\sigma$&  {U} &   {B}  \\
NIRSpec Prism & 19.16$\sigma$&  {U}  & 18.9$\sigma$ & {U} &  {B} & NIRSpec Prism &  16.15$\sigma$&  {U}  & 16.03$\sigma$&  {U} &   {B} \\
G235M + G395M  & 21.31$\sigma$&  {U} &21.31$\sigma$&  {U} &  {B} & G235M + G395M &   18.57$\sigma$ & {U}  & 18.64$\sigma$&  {U} &  {B}  \\
\hline
\end{tabular}
\end{center}
\caption{The results for all the test cases and instrument modes for the 3 retrieval styles used in this study. We calculated the Bayes factor of the 2-D and dilution style with respect to the 1-D approach (difference in the ln(Z) values) and then converted this to a significance estimate for the detection of more complex models in terms of standard deviation $\sigma$. We consider a strong detection if $>$3.6$\sigma$. The letters U and B indicate whether the retrieval was either biased or unbiased in terms of chemistry. All models had a reduced $\chi^2$ of below 1.}
\label{tab:80_results}
\end{table*}

\subsection{NIRSpec G140M}
This is the most narrow wavelength region (0.97 -- 1.84 $\mu$m) which is similar to the coverage of the Wide Field Camera 3 (WFC3) on the Hubble Space Telescope (1.1 -- 1.7 $\mu$m): hence, the main absorber in this region is H$_2$O (See Fig.~\ref{fig:molecules}). This instrument would therefore be ideal if repeat studies of objects observed with WFC3 were required at a higher spectral resolution. 

\begin{figure*}
    \centering
    \includegraphics[width=1.0\textwidth]{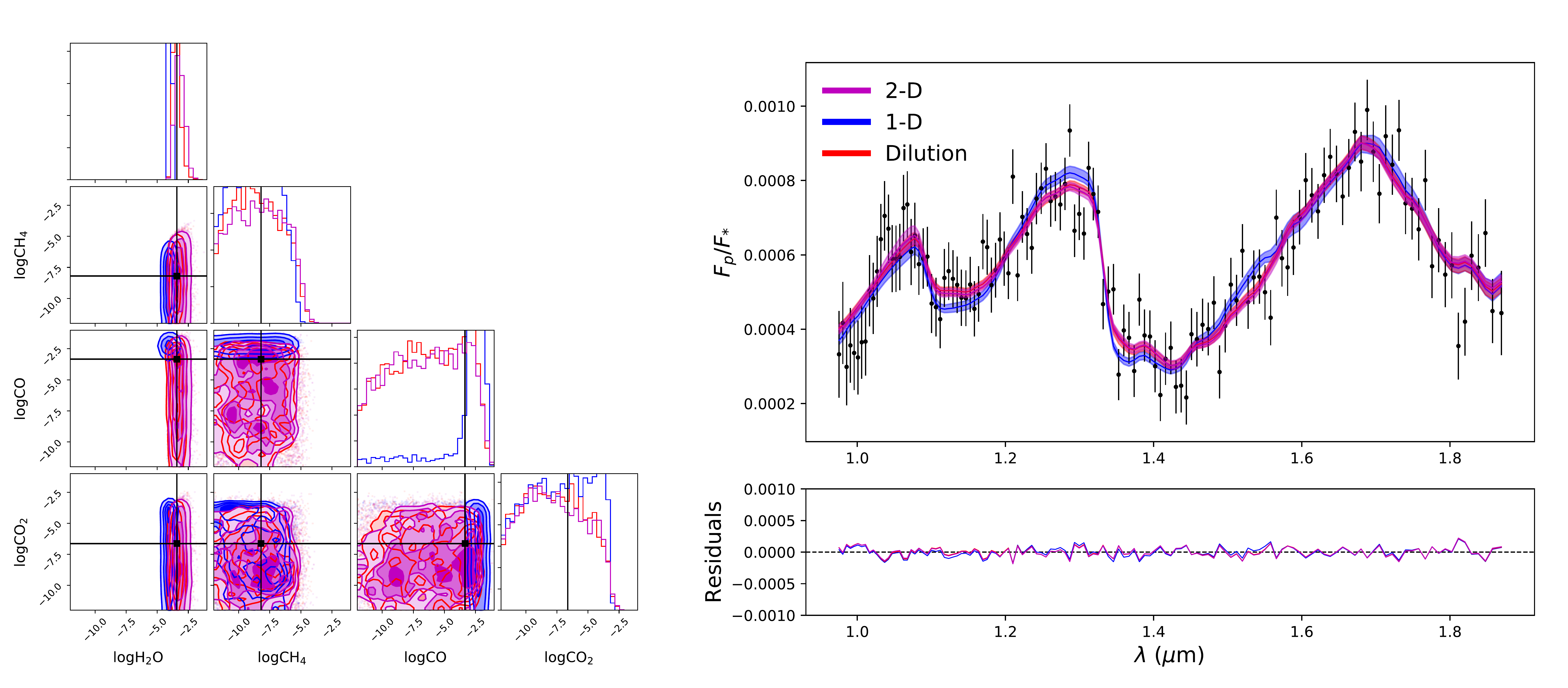}
    \caption{Best fitting models and accompanying posterior distributions for a planet scenario where the hot profile covers 60\% of the observed disk as seen using NIRSpec G140M. We overplot the three retrieval styles: 2-D (magenta), 1-D (blue) and Dilution (red) and show in the residuals (model-data) that they present good fits to the data.  The retrieved chemical abundances shown in the corner plot on the left are strongly biased compared to the true values (in black) for the 1-D retrieval case. This leads to a spurious detection of CO while no CO feature is seen in the original model (see the opacity comparison in Fig.~\ref{fig:molecules}).}
    \label{fig:60_G140}
\end{figure*}

With regards to the ability to disentangle the contribution from two profiles, Table \ref{tab:80_results} shows it has the capability to do so with a planet with 60\%, 40\% and 20\% of the flux contribution coming from the hot profile. For these simulations, the 1-D retrieval style produced biased chemistry results, whereas the 2-D and the dilution approaches did not. As the main absorber in this region is H$_2$O, an unbiased retrieval would be one where it was possible to constrain the abundance of this molecule and it was only possible to put an upper limit on the abundance of the other molecules. We present in Fig. \ref{fig:60_G140} a comparison for the 60\% scenario, where the 2-D and dilution retrievals were unbiased and the 1-D was biased.

In the $x$ = 80\% case, the H$_2$O abundance was constrained correctly for the 1-D retrieval. However, we also obtain a tight constraint on the CO$_2$ abundance at 3 orders of magnitude higher than the input value, corresponding to a spurious detection. For $x$ = 60\%, 40\% and 20\% the H$_2$O abundance from the 1-D retrieval was accurately constrained to smaller values than the input and CO was constrained to an order of magnitude larger than the input. Again this would be a spurious detection of CO as the absorption due to H$_2$O is roughly 1 order of magnitude higher than CO in this region (see Fig.~\ref{fig:molecules}).

For an atmosphere that has a flux contribution that is 80\% from the hotter profile, it is not possible to distinguish between 2-D or 1-D retrieval styles. This can be understood as the contribution from the colder profile, which already has muted spectral features, now has only a 20\% contribution to the final spectrum, which are lost in the noise. 

\subsection{NIRSpec G235M}
For all of the cases apart from $x$ = 20\% we see that only the 2-D or the dilution retrieval styles can correctly interpret the data. However, for the $x$ = 80\% case the significance of using 2-D or the dilution over a 1-D style would be considered as a non-detection; the chemistry in the 1-D style was also not biased. 

For all the unbiased scenarios, it was possible to constrain the abundance of H$_2$O and CO, but it was only possible to put upper limits on CH$_4$ and CO$_2$; this is to be expected as these molecules do not have strong features over this wavelength region (Fig. \ref{fig:molecules}).

Table \ref{tab:80_results} shows that for the $x$ = 20\% case, all three retrieval styles produced biased results. We performed another retrieval with the same model, this time without any noise added (but keeping the same uncertainties) and again we get biased chemistry. However, the posteriors were multimodal and one mode sampled the true chemistry values. We attribute this to SNR of the observation being too low that the retrieval is not able to pick out discerning features. We present in Fig. \ref{fig:60_G235} a comparison for the 60\% scenario, where the 2-D and dilution retrievals were unbiased and the 1-D retrieval was biased.



\subsection{NIRSpec G395M}
For all of the planet cases, it is not possible to detect the 2 TP profiles. In all cases apart from $x = 40$\%, if 2 TP profiles were not used in the retrieval, then biases in the chemistry arose. We ran the 40\% case with no noise model added and the results were then consistent with the other cases. Hence, the noise realisation used was causing the bias and not the SNR. The wavelength region covered by this instrument (2.87 -- 5.10 $\mu$m) has been deemed to be a crucial grating as it covers the molecular features of CO and CO$_2$ (See Fig.~\ref{fig:molecules}) and the photometry points of the Spitzer Space Telescope. By performing tests on this grating it has become apparent that long wavelength coverage by itself is not what is crucial, but it is the specific wavelength region we cover. This is highlighted from Fig.~\ref{fig:snr}, we can see that for the wavelengths covered by this grating, a SNR in the range of $\sim$10 would be needed to disentangle the 2-D effects in temperature, depending on the contrast of the profiles, hence we require shorter wavelengths to be able to disentangle the effects more efficiently. 

We want to emphasise that, despite there being no evidence to use a 2-D retrieval or dilution retrieval, if one is not use, the chemistry will tend to be biased. Thus, caution needs to be taken when interpreting Bayesian evidence.

We present in Fig. \ref{fig:60_G395} a comparison for the 60\% scenario, where the 2-D and dilution retrievals were unbiased and the 1-D retrieval was biased. 

\subsection{NIRSpec Prism}
\begin{figure*}
    \centering
    \includegraphics[scale=0.25]{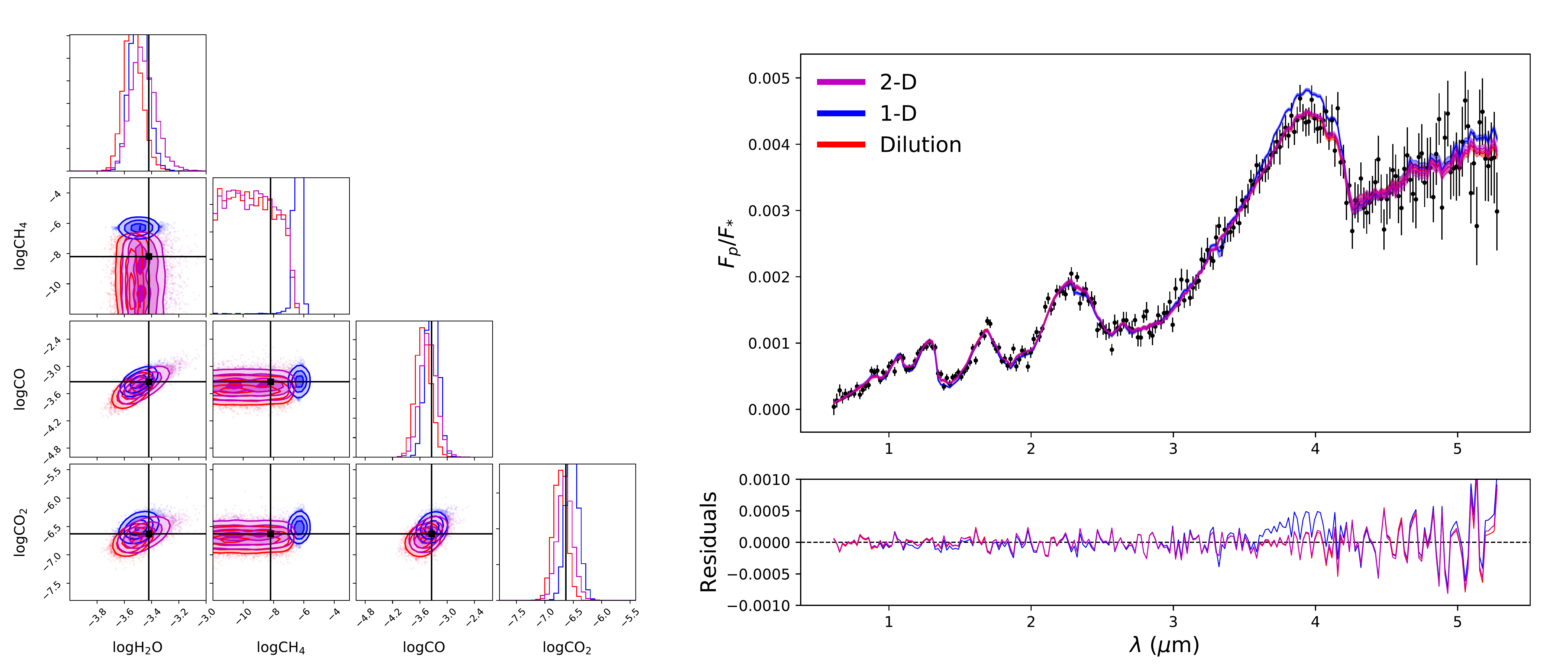}
    \includegraphics[scale=0.25]{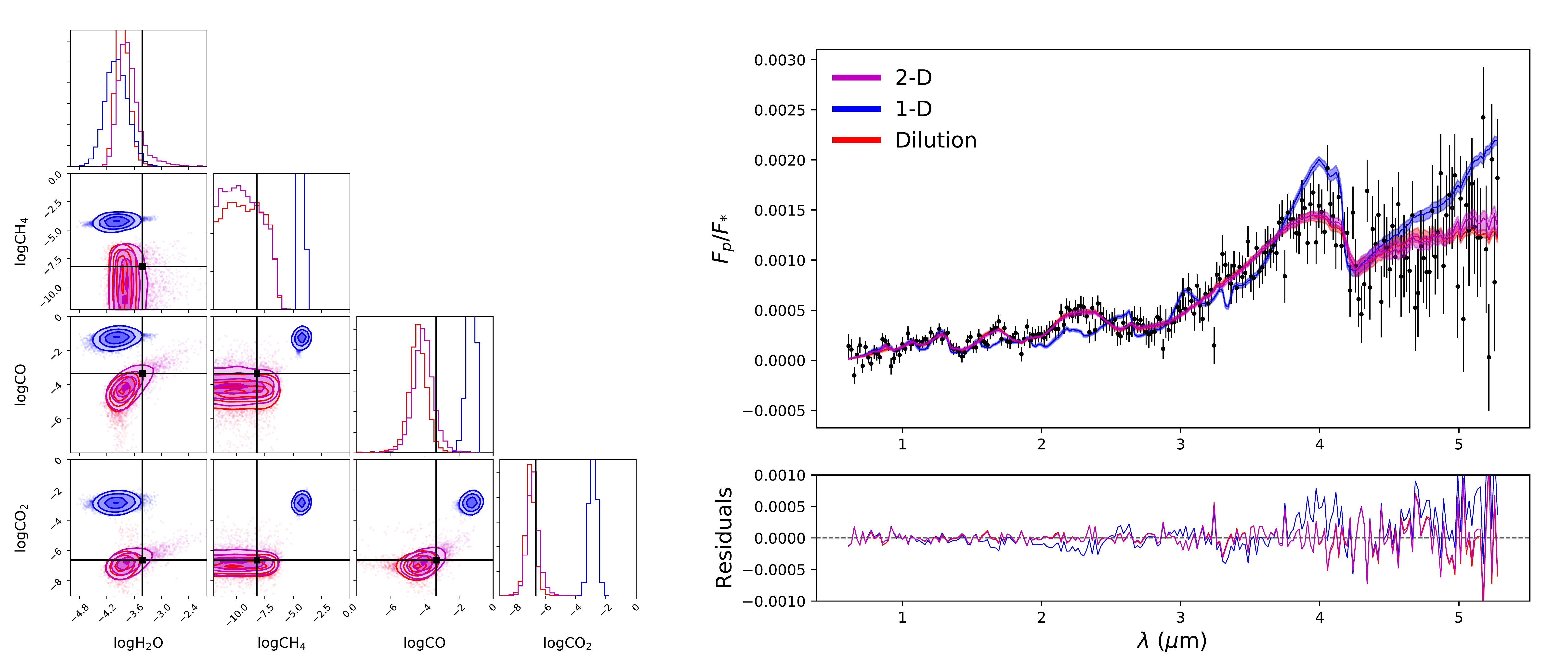}    
    \caption{Best fitting retrievals and accompanying posterior distributions for the planet scenarios with weightings $x$ = 80\% (top) and $x$ = 20\% (bottom), as seen using NIRSpec Prism. We overplot the three retrieval styles: 2-D (magenta), 1-D (blue) and Dilution (red) and show in the residuals (model-data) that they present good fits to the data. The true values are given by the solid black lines. When analysing the chemistry result presented in the triangle plots, we find a biasing of the result. Note that performing a 1-D retrieval presents a good fit to the data, however the chemistry is biased, only by exploring 2-D models and comparing the Bayesian evidence would this have been apparent. Hence caution needs to be taken when performing retrievals on data that is not inherently 1-D.}
    \label{fig:80_prism}
\end{figure*}

With the largest wavelength coverage of the study, we find significant evidence for using a 2-D or a dilution retrieval for each of the planetary cases. We show in Fig.~\ref{fig:80_prism} the best fitting models for each of the retrieval cases for both $x = 80$\% (top) and $x = 20$\% (bottom), the residuals for the fits and also the posteriors. It can be seen that each case provides a fit that has a reduced $\chi^2$ of below one; hence, without performing a model comparison, we could conclude that 1-D fit is good enough for the data. However, if we analyse the triangle plot, we see that for $x = 80$\%  the 1-D retrieval style produces a bias in the abundance and achieves a tight constraint on CH$_4$. For the $x = 20$\% case, the 1-D retrieval style biases all of the chemistry apart from the H$_2$O content.

\subsection{G235M + G395M}
By combining the two gratings we achieve a similar level of $\sigma$ confidence as using the Prism. What is interesting is that G395M alone was not able to disentangle the 2 TP profiles. The addition of the G295M grating meant that the spectral coverage was over a range that has a shorter wavelength, which is shown in Fig.~\ref{fig:snr} to be a crucial spectral region that needs to be covered. The required SNR increases as we move to longer wavelengths. We present in Fig. \ref{fig:60_G235+395} a comparison for the 60\% scenario, where the 2-D and dilution retrievals were unbiased and the 1-D was biased.

\subsection{2-D vs Dilution styles}
\label{2tpvdilution}
From Table~\ref{tab:80_results} it can be seen that the $\sigma$ significance of the 2-D and dilution retrieval styles compared to the 1-D style are comparable. The difference is insignificant and can be considered as a non-detection. We can interpret this to mean that fitting for a second TP profile for a planet which we know to have one, is not favoured given the resolution of the data. Instead, a dilution factor applied to a 1-D retrieval style is all we can interpret. The dilution factor mimics the weight that is applied to the hotter profile; for example, the dilution factor obtained from NIRSpec Prism for $x$ = 0.8, 0.6, 0.4 and 0.2 are $s$ = 0.84, 0.67, 0.51, and 0.32 respectively. 

It can be seen from Table~\ref{tab:80_results} that the performance of the chemistry retrieval for both styles is comparable. Out of the 20 retrievals performed with each technique, both the 2-D and dilution retrievals produced biased chemistry in 2/20 of the cases. This is by far an improvement over the 1-D style which produced biased chemistry in 18/20 of the retrievals.

\subsection{Origins of the chemistry bias}
\label{sec:Bias}
As we have seen in Section \ref{section4} and as discussed in \citet{feng2016impact}, in many cases performing a 1-D retrieval on a model created with 2 TP profiles can lead to biased retrieved abundances and even the spurious detection of chemical species. We now try to understand which mechanism can produce this bias.

To first order, we can assume that a planets atmosphere emits at a given wavelength with an amount of flux equivalent to the Planck function at the local temperature for this wavelength. For a planet with a temperature that decreases with decreasing pressure (i.e. a non-inverted temperature-pressure profile), the thermal emission is larger if there are less absorbers in this region, and smaller if there are more absorbers in this region.

An emission spectrum at the resolution of JWST contains information at different scales. At small spectral scales, the molecular features corresponding to single molecules are detected and their slope puts a strong constraint on the slope of thermal profile, with each band constraining the temperature gradient at a different layer. We see from our simulations the retrieval uses the numerous H$_2$O bands to determine the thermal structure of the atmosphere. However, as seen in the bottom panel of Fig. \ref{fig:bb_cross}, the presence of an inhomogeneous atmosphere should affect the general slope of the spectrum (e.g. the mean slope between 1 and 5 microns, hence a larger spectral scale). Such a slope could be interpreted by the retrieval as a change in the slope of the temperature-pressure profile, however, because the temperature gradient is set in each layer by the small spectral scales (e.g. the slope of molecular features), the model cannot change the temperature-pressure profile to adjust the general slope of the spectrum. It must therefore rely on varying the opacities, hence the abundances. 


Now let us imagine that we try to retrieve the abundances and temperature profile of a planet with a diluted hot spot, i.e., a hot temperature profile on a contained part of the visible hemisphere of the planet. At a given layer, the diluted blackbody would be given by the blue curve in Fig.~\ref{fig:bb_cross}. When trying to retrieve the planetary spectrum with a 1-D model, a typical blackbody would look like the red curve of Fig.~\ref{fig:bb_cross}. The ratio between the two, shown in the bottom panel of Fig.~\ref{fig:bb_cross}, have a slope: the 1-D emission is always smaller than the diluted 2-D emission at short wavelengths and larger at large wavelengths. 

\begin{figure*}
    \centering
    \includegraphics[scale=0.27]{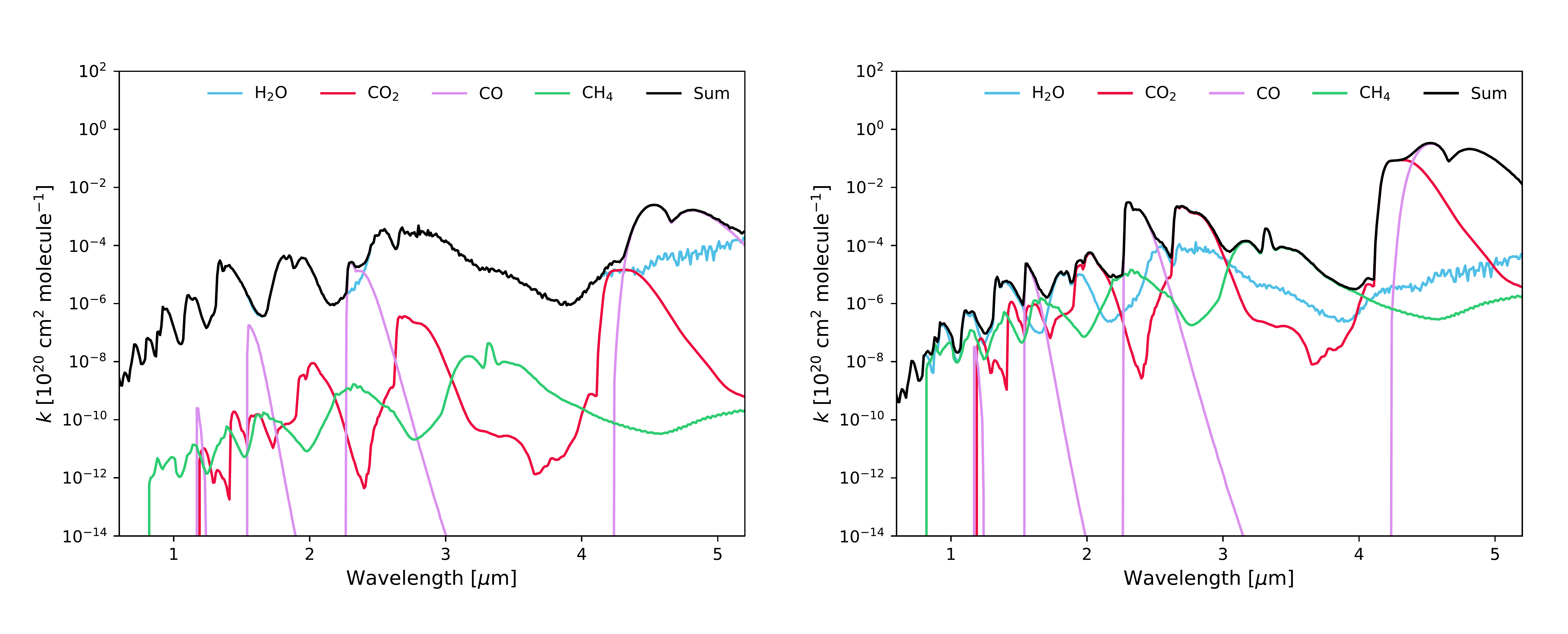}
    \caption{\textbf{Left:} Weighted-by-abundance absorption coefficients of the molecules from the best fit 2-D model for the 20\% hot flux case shown in Fig.~\ref{fig:80_prism}; we also present the sum of the absorption coefficients, which is representative of the absorption of the final best fitting spectrum. The sum is dominated by the water opacity at most wavelengths. \textbf{Right:} Weighted-by-abundance absorption coefficients of the molecules from the best fit 1-D model for the 20\% hot flux case shown in Fig.~\ref{fig:80_prism}; we also present the sum of the absorption coefficients, which is representative of the absorption of the final best fitting spectrum. We see that the 1-D retrieval requires an increase in the opacity of all the molecules apart from H$_2$O to fit to the observations. This is because varying H$_2$O would shift the total opacity up and down without changing the slope of the opacities at large spectral scales. Other molecules, such as CH$_4$ or CO have a larger variation between 1 and 5 microns and are used by the retrieval to increase the mean slope of the total atmospheric opacities.}
    \label{fig:molecule_compare}
\end{figure*}

One could adjust this slope in two possible ways, the first is to vary the vertical temperature gradient, so that different layers are probed at different wavelengths. However, the temperature gradient is already determined by the small scale molecular spectral bands that are detected. The other way, is to change the levels that are probed, i.e. to change the opacities so that the long wavelengths probe cooler layers and the short wavelengths probe hotter layers. For a non-inverted temperature-pressure profile that means that the opacity should be increased at long wavelengths and decreased at short wavelengths. As seen in the left panel of Fig.~\ref{fig:molecule_compare}, H$_2$O is dominating the diluted hot spot spectrum. Hence increasing the H$_2$O abundance cannot significantly change the large scale spectral slope of the opacities. As seen in the right panel of Fig.~\ref{fig:molecule_compare}, to change the spectral slope of the opacities we can increase the abundances of species that have a larger spectral slope than H$_2$O, e.g., CH$_4$ and CO. The combination of CH$_4$ and CO opacity can mimic the small scale H$_2$O features while increasing the general spectral slope of the opacities. We note that NH$_3$, considered in \citet{feng2016impact} but not in this study, could play a similar role as CH$_4$ and CO, giving even more opportunities to mimic the small scale signatures of H$_2$O while giving more liberty to adjust the large scale spectral slope of the opacities. More generally, any additional molecule would give an additional leverage to adjust the spectral slope and compensate the lack of 2 TP profiles in the original fit. This would then lead to spurious molecule detection when using a model with 1 TP profile. Given that any additional molecule considered in a retrieval framework would increase the ability of the retrieval to compensate for the lack of inhomogeneities, this has the potential to lead to numerous spurious detection of molecules in JWST spectra. We therefore urge future modellers to carefully quantify the possible effects of temperature inhomogeneities in a planet atmosphere before claiming the detection of an unexpected molecule.

\subsection{Wavelength dependent radius effect}
\begin{figure}
    \centering
    \includegraphics[scale=0.45]{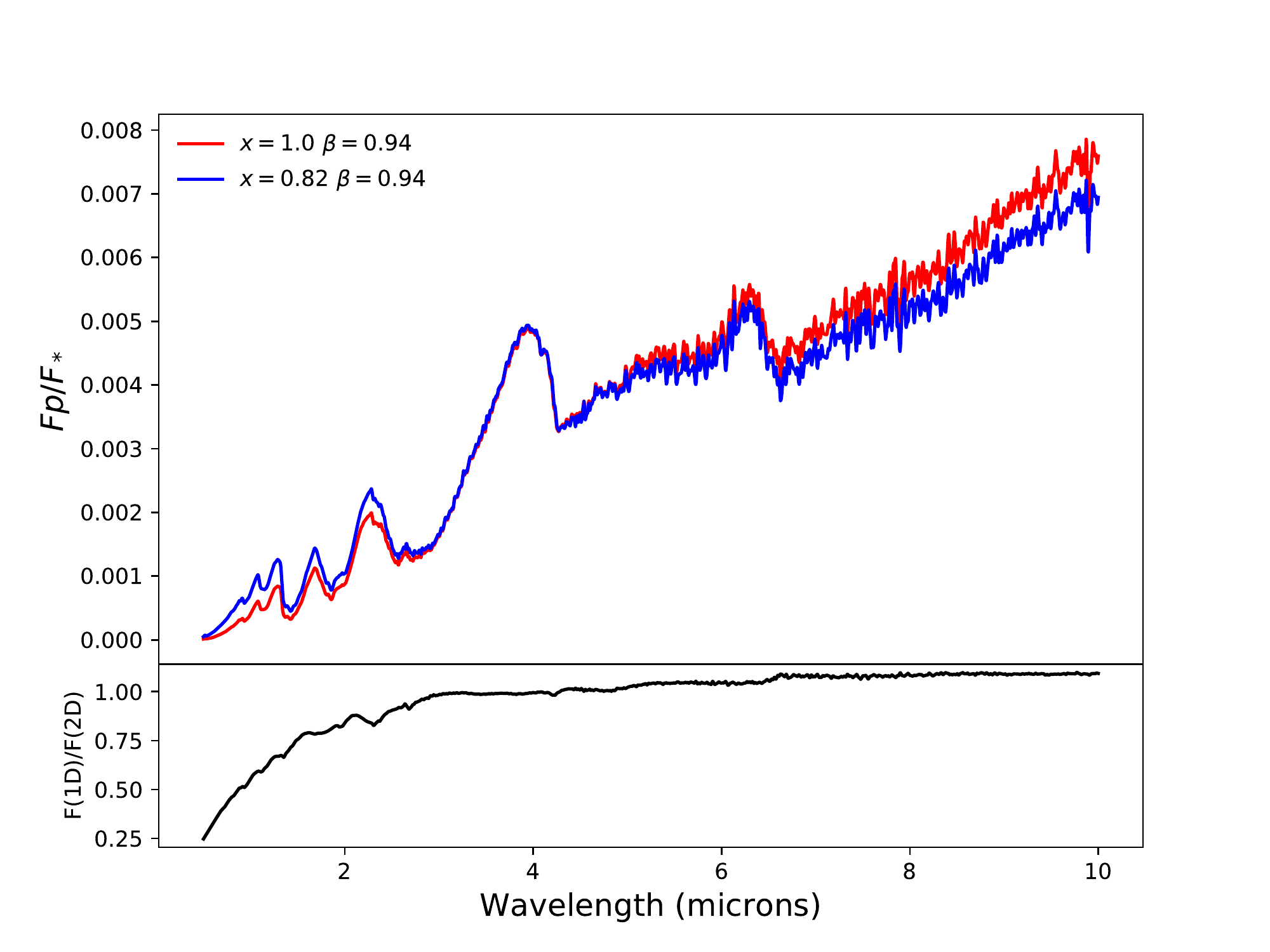}
    \caption{Top: Dilution and 1-D forward models generated from the best fitting parameters from the x = 80\% case for the G395M observing mode. For the 1-D, $x = 1.0$ and $\beta = 0.94$ were retrieved and used for the forward model, for the dilution, which is analogous for 2-D, $x = 0.82$ and $\beta = 0.94$ were retrieved and used in the forward model. We chose to compare these due to their chemistry being unbiased in both scenarios, hence we are seeing the effect due to the thermal profiles. Bottom: The ratio of the two models. It can be seen that there is a larger gradient between the two at shorter wavelengths and this begins to plateau at longer wavelengths, this is similar to our analytical model in Fig \ref{fig:bb_cross}. }
    \label{fig:ratio}
\end{figure}

Recent investigation into other biases that arise when studying emission spectra include the wavelength dependence of the planet radius \citep{drummond2018effect,gandhi2018retrieval,Fortney2019}. \citet{Fortney2019} highlight that this effect is not an offset, and can cause a planet-to-star flux ratio variation of 5\% for a typical hot Jupiter. Our results would remain unchanged when modelling the wavelength dependent radius effect, this is because the biases we find arise from the gradient of the opacities observed over a large wavelength range (i.e., a large spectral scale), whereas the radius effect would have its largest effect on the smaller scales of the spectrum, particularly by changing the shape of each molecular absorption bands. We demonstrate this in Fig \ref{fig:ratio} by showing the ratio of the 2-D and 1-D best fitting models from the x = 80\% G395M simulations which had unbiased chemistry results. We see that there do not appear to be many small scale fluctuations in the residuals (bottom panel of Fig. \ref{fig:ratio}, compared to bottom panel of Fig. 2 from \citet{Fortney2019}) hence the wavelength dependent radius effect and bias due to 2-D structures are independent of one and other, and given good enough SNR, it should be possible to disentangle both of them. 

We also would like to note that modelling the wavelength dependent radius effect in 1-D would probably lead to the over estimation of the effect for the dayside spectrum of a planet. The effect of the variation of the radius is proportional to the scale height at the limb and is therefore sensitive to the limb temperature. However, the spectrum is sensitive to the temperature of the planets dayside, which is usually hotter than the limb temperature. By assuming that the correction due to the radius effect must be done at the same temperature as the retrieved temperature from the spectrum, one would likely overestimate the importance of the radius effect. One possible way to model the radius effect would be to model it in a 2-D framework like the one this study presents. When analysing a dayside spectrum, the transit radius variation should be determined by the retrieved cold temperature. For phase curve observations, the viewing angle of the limb will vary depending on the phase of the observation, hence $\Delta R_p(\lambda) = |yR_p(T_{\text{hot}}) + (y-1)R_p(T_{\text{cold}})|$ where y would be a parameter which would represent the weighting of the limb seen by the observer. 


\section{Implication for Future Space Observations}
\label{section5}
\subsection{Application to ERS observations}
The first emission spectra observations of WASP-43b using the James Webb Space Telescope will be be conducted from the Transiting Exoplanet Community Early Release Science Program \citep{bean2018transiting}. They will obtain a phase curve observation using MIRI/LRS that will consist of two secondary eclipse observations and one primary transit observation. The wavelength grid covered (5 -- 12 $\mu$m) and resolution (R $\sim$ 100) are similar to those investigated in this study, i.e NIRSpec Prism and combining the gratings G235M and G395M. However, the wavelength coverage does not go short enough to be able to disentangle the temperature of the 2 profiles. Fig. \ref{fig:snr} shows that potentially a SNR in the range 10 -- 100 will be needed.

\subsection{Application to GTO observations}
There is a planned phase curve observation of WASP-43b in the JWST Guaranteed Time Observations (GTO) program (PI: Stephen Birkmann, Proposal 1224). They propose to use NIRSpec G395H, which is the higher resolution grating. We show that studying the horizontal temperature structure of this planet is difficult with this grating and caution needs to be taken when thinking about interpretation of the data set. In no cases do our results show that a 2-D retrieval is favoured over a 1-D retrieval, this is attributed to the wavelength coverage only sampling the tail end of the blackbody function. We do show that in  three out of the four case, the chemistry result is biased if a 2-D or dilution approach is not used.

For a planetary case that had 40\% flux coming from the hot profile, all of the cases produced a bias result in the chemistry. When analysing without a noise model, the 2-D and dilution retrieval presented unbiased results.

Analysis of these data alone should be carefully interpreted: the results may vary when combined with the ERS observations to cover more molecular features.

\section{Conclusions}
\label{section6}
 
We investigate how an inhomogeneous horizontal temperature structure effects the way we perform emission retrievals. We aim to understand the origins of the biases and how to mitigate them with JWST's NIRSpec instrument. 

We first derive an analytical criterion which can be used to find the signal-to-noise ratio required to disentangle inhomogeneous temperature structures. We find that the signal-to-noise ratio needed increases for longer wavelengths, this is demonstrated in Fig. \ref{fig:snr}. This is because the longer the wavelength, the closer the emission spectra from the hot and cold parts of the atmosphere are to each other.

We then build planetary scenarios with varying hot and cold flux contributions. We generate synthetic observations for the NIRSpec observing modes and perform atmospheric retrievals on them. We find that performing a homogeneous retrieval will often result in biased chemical abundances, as shown from Table \ref{tab:80_results}.

We show that a novel method of applying a dilution factor to a 1-D model can provide a similar retrieval capability as using the 2-D technique, for cases where the hot-spot is significantly hotter than the surrounding atmosphere, suggesting that it might be an ideal technique for studying the chemistry of the atmosphere while saving on computation time. This is because it has the capability to perform the same as either a 1-D retrieval (if the dilution = 1) or a 2-D retrieval (if dilution $\neq$ 1) without facing biases in the chemistry.


Due to the large amount of dedicated time needed to perform phase curve observations, it is prudent to know which instruments are necessary to study the 2-D structure in more detail. We find that with a large enough spectral coverage we can break the correlation between the amount of flux output and the temperature for all of the planetary cases explored in this study (i.e using Prism or combining G235M and G395M). Crucially, we show that the shorter the wavelength observed the better the detection of 2-D effects. We find that, despite G395M being the grating with the largest spectral coverage, there was no evidence to support using a 2-D or dilution retrieval over a 1-D style, but if these retrieval styles were not used, then biased results were produced. This reiterates that not only do we need a large wavelength coverage, but this coverage needs to sample shorter wavelengths. As ARIEL \citep{tinetti2018chemical} has a large wavelength coverage of 0.5 - 7.8 $\mu$m, it is ideal for detecting and quantifying inhomogeneities.

Smaller spectral coverage can also break the degeneracy, such as using G140M (sampling before the peak of the blackbody function), although with smaller significance and only if the hot profile contributes 60\% or less to the final spectrum. This is something that would not be known for the dayside of a planet unless performing \textit{a piori} studies with GCM models. It would, however, be known for tidally-locked planets that are observed at various points in its phase around the star. 

Finally, we explain the origins of the biases that we face when performing 1-D retrievals on a planetary system that has an inhomogeneous temperature structure. A 1-D model matching an emission spectrum from an inhomogeneous planet underestimates the spectrum at short wavelength and overestimate it at long wavelengths. As a consequence, the 1-D model requires larger opacities at longer wavelengths and smaller opacities at shorter wavelengths. In other words, the mean slope of the opacity vs. wavelength needs to be larger in the 1-D model than in the original spectrum. When H$_2$O dominates the original spectrum, this naturally leads to an increase in the abundance of molecules that have an opacity that increases more sharply with wavelength than H$_2$O, e.g. CH$_4$ or CO. Importantly, the larger the number of molecules considered, the easier it is for the 1-D retrieval to find a way to compensate for its lack of inhomogeneities. We therefore recommend that future works should consider the possibility of temperature inhomogeneities when searching for new molecules in JWST data sets in order to ensure that these detections are not spurious.



\section*{Acknowledgements}

J. Taylor is a Penrose Graduate Scholar and would like to thank the Oxford Physics Endowment for Graduates (OXPEG) for funding this research. P.G.J. Irwin acknowledges the support of the United Kingdom's Science and Technology Facilities Council. We would like to thank the referee for their kind and thorough comments which have helped to improve this manuscript.




\bibliographystyle{mnras}
\bibliography{bibliography} 



\appendix

\section{Complete Analytical Formulation}
Here we highlight the intermediate steps between Eq \ref{flux_soln} and Eq \ref{criterion}. We want to show how the flux varies with wavelength and then by $\beta$, first we differentiate Eq \ref{flux_soln} to get

\begin{dmath}
\label{dqdw}
\frac{\delta F(\lambda,\beta,x)}{\delta\lambda} = \frac{2c^3h^2e^{\frac{ch}{k \beta T \lambda}}\Big(-1 + e^{\frac{ch}{\beta T \lambda_0 k}}\Big) x_0}{\Big(-1 + e^{\frac{ch}{\beta_0 T \lambda_0 k}}\Big)\Big(-1 + e^{\frac{ch}{\beta T \lambda k}}\Big)^2k\beta T \lambda^7} - \frac{10c^2h\Big(-1 + e^{\frac{ch}{\beta T \lambda_0 k}}\Big)x_0}{\Big(-1 + e^{\frac{ch}{\beta_0 T \lambda_0 k}}\Big)\Big(-1 + e^{\frac{ch}{\beta T \lambda k}}\Big)\lambda^6}
\end{dmath}

Finally, by differentiating with respect to $\beta$ we find the relationship between the given wavelength and how hot the region of the planet under consideration is. 
\begin{dmath}
\label{dqdwdb}
    \frac{\delta}{\delta\beta}\Big(\frac{\delta F(\lambda,\beta,x)}{\delta\lambda}\Big) = \frac{4 c^4h^3 e^{\frac{2 ch}{\beta T \lambda k}} \left(-1+e^{\frac{ch}{k \lambda_0 \beta T}}\right) x_0}{\left(-1+e^{\frac{ch}{k \beta_0 \lambda_0 T}}\right) \left(-1+e^{\frac{ch}{k \beta T \lambda}}\right)^3 k^2 \beta^3 T^2 \lambda^8}-\frac{2 c^4h^3 e^{\frac{ch}{k \beta T \lambda}} \left(-1+e^{\frac{ch}{k \lambda_0 \beta T}}\right) x_0}{\left(-1+e^{\frac{ch}{k \beta_0 \lambda_0 T}}\right) \left(-1+e^{\frac{ch}{k \beta T \lambda}}\right)^2 k^2 \beta^3 T^2 \lambda^8}-\frac{2 c^4h^3 e^{\frac{ch}{k \lambda_0 \beta T}+\frac{ch}{k \beta T \lambda}} x_0}{\left(-1+e^{\frac{ch}{k \beta_0 \lambda_0 T}}\right) \left(-1+e^{\frac{ch}{k \beta T \lambda}}\right)^2 k^2 \lambda_0 \beta^3 T^2 \lambda^7}-\frac{12 c^3h^2 e^{\frac{ch}{k \beta T \lambda}} \left(-1+e^{\frac{ch}{k \lambda_0 \beta T}}\right) x_0}{\left(-1+e^{\frac{ch}{k \beta_0 \lambda_0 T}}\right) \left(-1+e^{\frac{ch}{k \beta T \lambda}}\right)^2 k \beta^2 T \lambda^7}+\frac{10 c^3h^2 e^{\frac{ch}{k \lambda_0 \beta T}} x_0}{\left(-1+e^{\frac{ch}{k \beta_0 \lambda_0 T}}\right) \left(-1+e^{\frac{ch}{k \beta T \lambda}}\right) \lambda_0 k \beta^2 T \lambda^6}
\end{dmath}

We can then use this to set a criterion for the detectability of this relationship as shown in Eq~\ref{criterion}.

\section{Extra Figures}

\begin{figure*}
    \centering
    \includegraphics[width=1.0\textwidth]{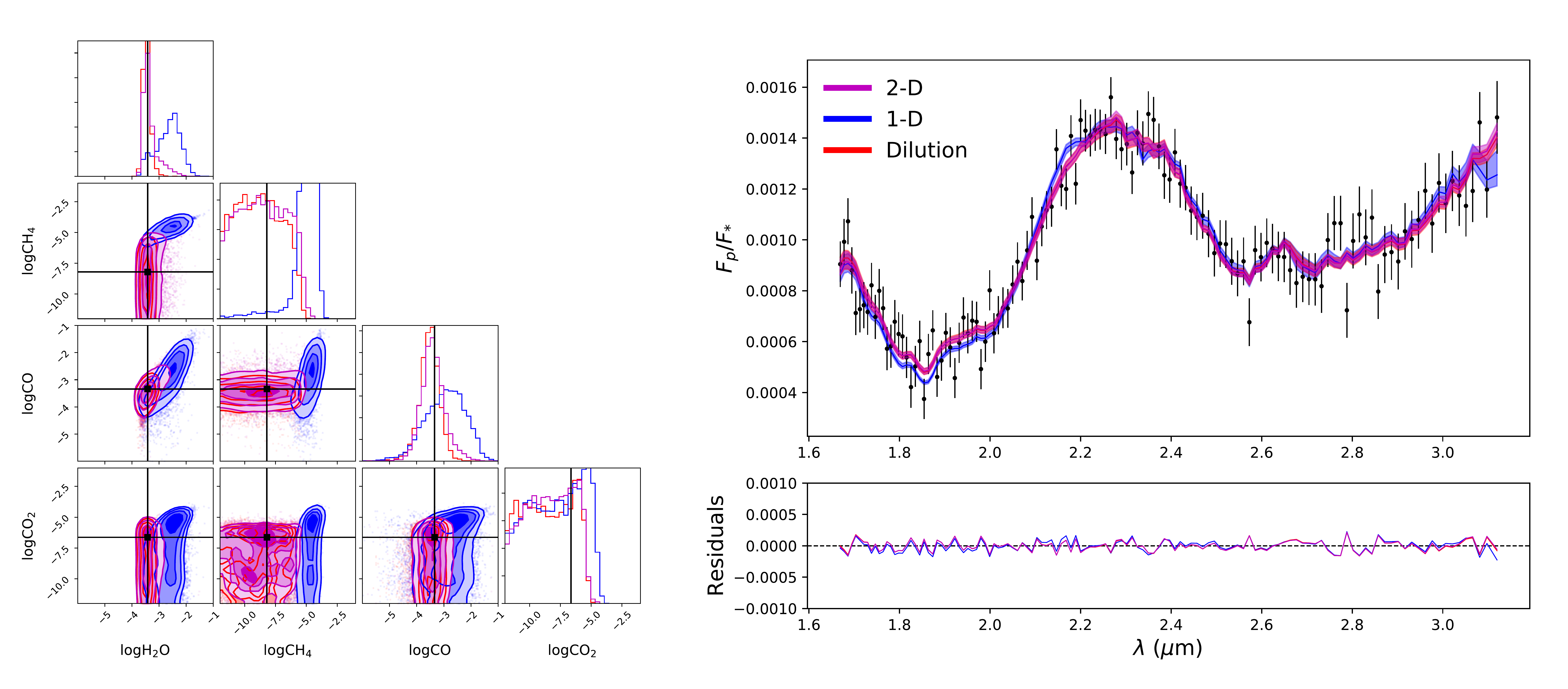}
    \caption{Best fitting retrievals and accompanying posterior distributions for a planet scenario of 60\% flux applied to the hot profile as seen using NIRSpec G235M. We overplot the three retrieval styles: We overplot the three retrieval styles: 2-D (magenta), 1-D (blue) and Dilution (red) and show in the residuals (model-data) that they present good fits to the data. The true values are given by the solid black lines. When analysing the chemistry result presented in the triangle plot, we find a biasing of the result from the 1-D style, we see that the 1-D retrieval detects H$_2$O and CH$_4$ at values which are outside 2-$\sigma$ from the input.}
    \label{fig:60_G235}
\end{figure*}

\begin{figure*}
    \centering
    \includegraphics[width=1.0\textwidth]{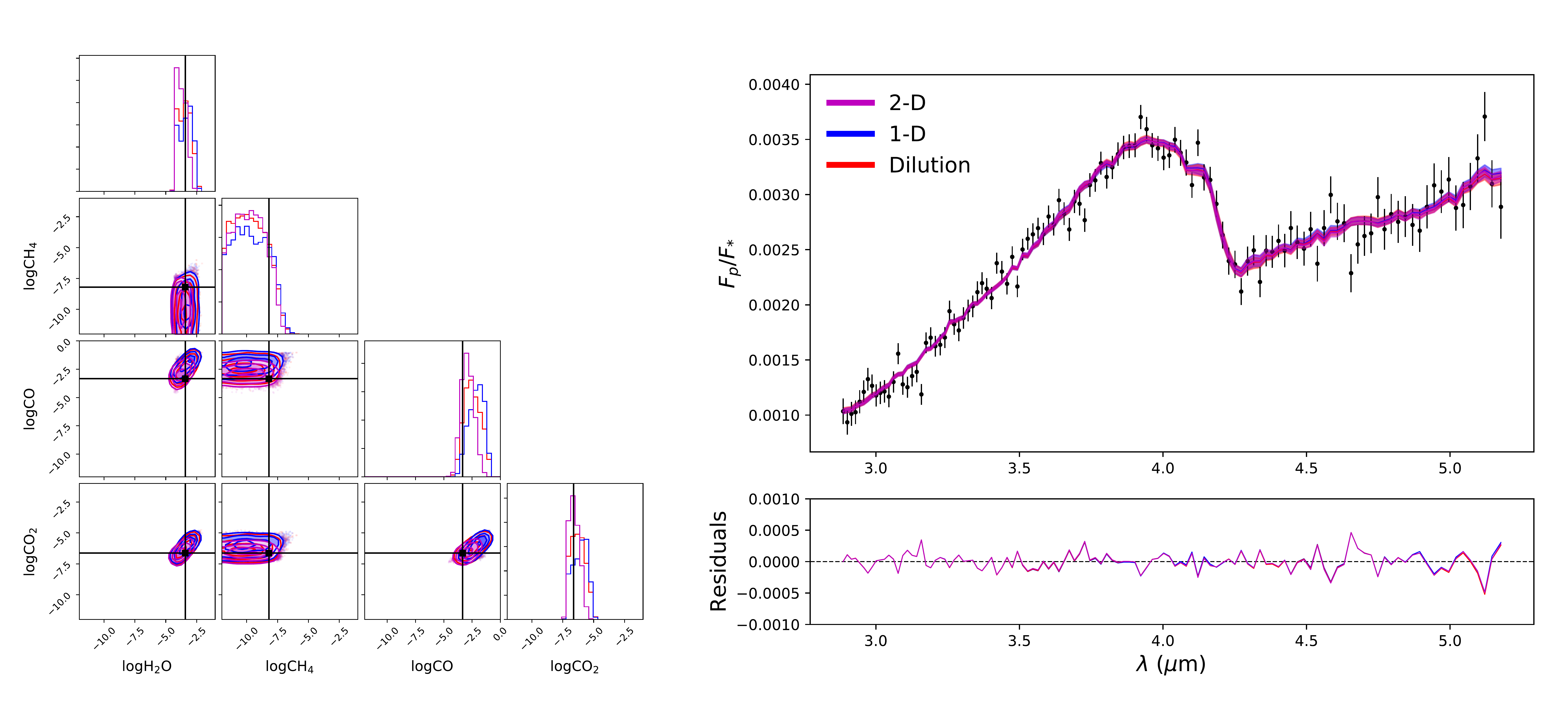}
    \caption{Best fitting retrievals and accompanying posterior distributions for a planet scenario of 60\% flux applied to the hot profile as seen using NIRSpec G395M. We overplot the three retrieval styles: 2-D (magenta), 1-D (blue) and Dilution (red) and show in the residuals (model-data) that they present good fits to the data. The true values are given by the solid black lines. When analysing the chemistry result presented in the triangle plot, we find a biasing of the result from the 1-D style, we see that the 1-D retrieval detects CO at a value which is outside 2-$\sigma$ from the input.}
    \label{fig:60_G395}
\end{figure*}


\begin{figure*}
    \centering
    \includegraphics[width=1.0\textwidth]{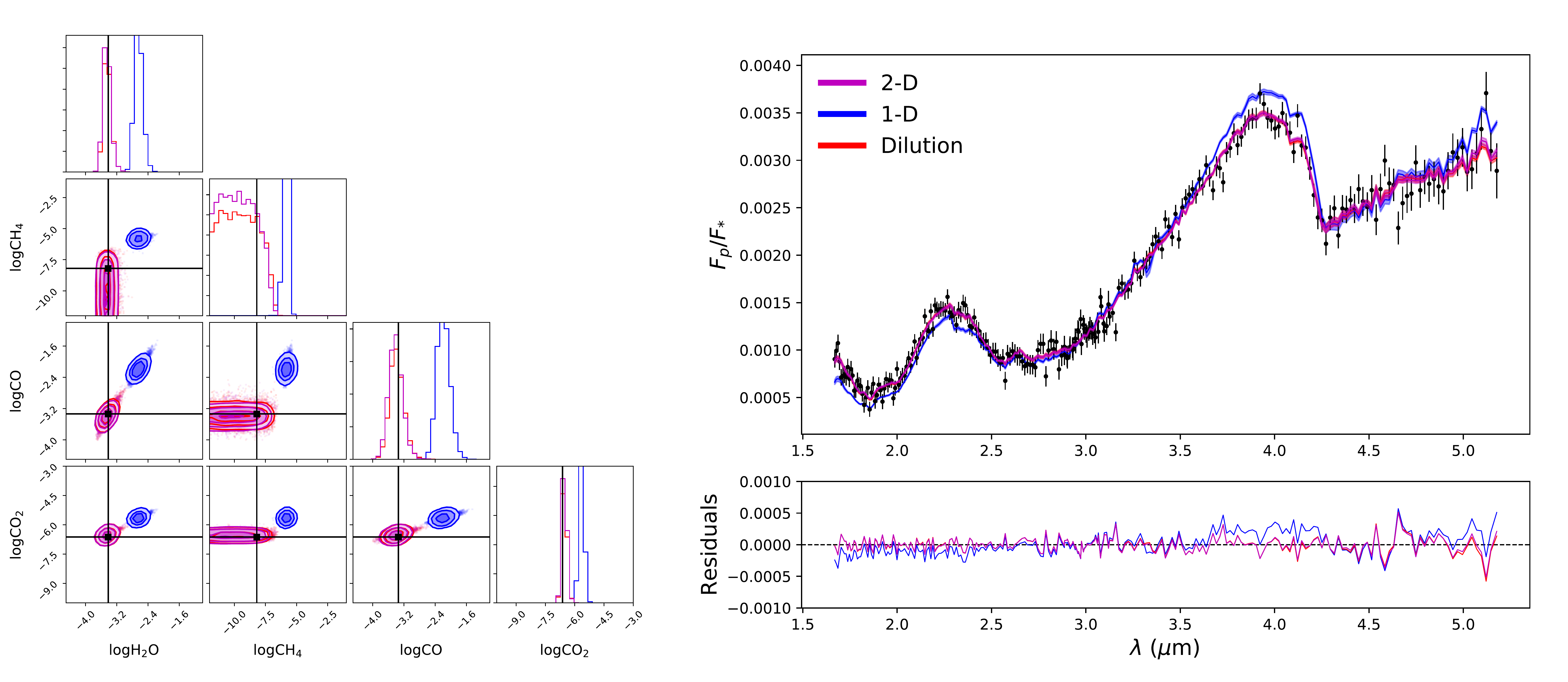}
    \caption{Best fitting retrievals and accompanying posterior distributions for a planet scenario of 60\% flux applied to the hot profile as seen by combining NIRSpec G235M and G395M. We overplot the three retrieval styles: 2-D (magenta), 1-D (blue) and Dilution (red) and show in the residuals (model-data) that they present good fits to the data. The true values are given by the solid black lines. When analysing the chemistry result presented in the triangle plot, we find a biasing of the result from the 1-D style, we see that the 1-D retrieval detects H$_2$O, CH$_4$, CO and CO$_2$ at values which are outside 2-$\sigma$ from the input.}
    \label{fig:60_G235+395}
\end{figure*}


\bsp	
\label{lastpage}
\end{document}